\documentclass[prd, 
showpacs, floatfix, nofootinbib,
preprintnumbers
]{revtex4}
\usepackage{amsmath,amsfonts,mathtools}
\usepackage{amssymb}
\usepackage{epsfig}
\usepackage{graphicx}
\usepackage{color}
\usepackage{hyperref}
\definecolor{darkred}{RGB}{150,0,0}
\definecolor{darkblue}{RGB}{0,0,150}
\hypersetup{colorlinks=true,linkcolor=darkblue,citecolor=darkblue,urlcolor=black}
\usepackage{comment}
\usepackage{bbold}
\usepackage{booktabs}
\usepackage[normalem]{ulem}
\usepackage{vwcol} 
\usepackage{multirow,bigdelim,dcolumn}

\newcommand{\be}{\begin{equation}}
\newcommand{\ee}{\end{equation}}
\newcommand{\bdm}{\begin{displaymath}}
\newcommand{\edm}{\end{displaymath}}
\newcommand{\bea}{\begin{eqnarray}}
\newcommand{\eea}{\end{eqnarray}}
\newcommand{\bpm}{\begin{pmatrix}}
\newcommand{\epm}{\end{pmatrix}}
\newcommand\lsim{\mathrel{\rlap{\lower4pt\hbox{\hskip1pt$\sim$}}
    \raise1pt\hbox{$<$}}}
\newcommand\gsim{\mathrel{\rlap{\lower4pt\hbox{\hskip1pt$\sim$}}
    \raise1pt\hbox{$>$}}}
\DeclareMathOperator{\Tr}{Tr}
\newcommand{\nn}{\nonumber}
\newcommand{\noi}{\noindent}
\newcommand{\RED}{\textcolor{darkred}}

\newcommand{\GeV}{\, \mathrm{GeV}}

\newcommand{\LR}{$SU(3)_c \times \protect\linebreak[0]SU(2)_L \times \protect\linebreak[0]SU(2)_R \times \protect\linebreak[0]U(1)_{B-L} \ $}
\newcommand{\PS}{$SU(4)_c \times \protect\linebreak[0]SU(2)_L \times \protect\linebreak[0]SU(2)_R \ $}

\newcommand{\vev}[1]{\langle #1 \rangle}

\def\TABSPACE{0.2em}
\def\PAR{\textcolor{black}}
\def\PARtm{\textcolor{black}}
\def\IND{\textcolor{blue}}
\def\Separator{;}
\def\Separatorb{,}

\AtBeginDocument{
\heavyrulewidth=.08em
\lightrulewidth=.05em
\cmidrulewidth=.03em
\belowrulesep=.65ex
\belowbottomsep=0pt
\aboverulesep=.4ex
\abovetopsep=0pt
\cmidrulesep=\doublerulesep
\cmidrulekern=.5em
\defaultaddspace=.5em
}
\begin{document}

\title{One-loop pseudo-Goldstone masses in the minimal  $SO(10)$ Higgs model}
\preprint{CETUP2016-005}
\author{Luk\'a\v{s} Gr\'af}\email{lukas.graf.14@ucl.ac.uk}
\affiliation{
Department of Physics \& Astronomy, University College London, Gower Street,
WC1E 6BT London, United Kingdom
}
\author{Michal Malinsk\'{y}}\email{malinsky@ipnp.troja.mff.cuni.cz}
\affiliation{Institute of Particle and Nuclear Physics,
Faculty of Mathematics and Physics,
Charles University in Prague, V Hole\v{s}ovi\v{c}k\'ach 2,
180 00 Praha 8, Czech Republic}
\author{Timon Mede}\email{tmede@phy.hr}
\affiliation{
Department of Physics, Faculty of Science, University of Zagreb, Bijeni\v{c}ka cesta 32,
HR-10000 Zagreb, Croatia
}
\author{Vasja Susi\v c}\email{vasja.susic@unibas.ch}
\affiliation{
Department of Physics, University of Basel,
Klingelbergstrasse 82, CH-4056 Basel, Switzerland
}
\begin{abstract}
We calculate the prominent perturbative contributions shaping the one-loop scalar spectrum of the minimal non-supersymmetric renormalizable $SO(10)$ Higgs model whose unified gauge symmetry is spontaneously broken by an adjoint scalar. Focusing on its potentially realistic $45\oplus 126$ variant in which the rank is reduced by a VEV of the 5-index self-dual antisymmetric tensor, we provide a thorough analysis of the corresponding one-loop Coleman-Weinberg potential, paying particular attention to the masses of the potentially tachyonic  pseudo-Goldstone bosons (PGBs) transforming as $(8,1,0)$ and $(1,3,0)$ under the Standard Model gauge group. The results confirm the assumed existence of extended regions in the parameter space supporting a locally stable SM-like quantum vacuum inaccessible at the tree-level. The effective potential (EP) tedium is compared to that encountered in the previously studied $45\oplus 16$ $SO(10)$ Higgs model where the polynomial corrections to the relevant pseudo-Goldstone masses turn out to be easily calculable within a very simplified purely diagrammatic approach.
\end{abstract}
\maketitle
\tableofcontents
\section{Introduction\label{sect:Introduction}}
The upcoming generation of very large volume detectors such as Hyper-K~\cite{Abe:2011ts} and/or DUNE~\cite{Adams:2013qkq} is not only a blessing for the neutrino community but it is also likely to provide a great deal of information to other branches of  particle physics research. Concerning, in particular, the possible baryon number non-conservation signals such as proton decay, the sensitivity of the current searches may be improved by as much as one order of magnitude, reaching up to about $10^{35}$~years for the proton lifetime.

Unfortunately, this steady progress is not matched by any significant developments on the theory side.  As a matter of fact, the existing proton lifetime estimates -- usually made in the context of grand unified theories (GUTs)~\cite{Georgi:1974sy}, the most economical scheme for addressing these issues in the standard quantum field theory context -- are typically plagued by theoretical uncertainties stretching over many orders of magnitude, see, for instance, Table II in~\cite{Kolesova:2014jua} and references therein. Needless to say, this is way too poor to make any real benefit from the expected experimental sensitivity improvements (unless we were lucky and a clear signal of baryon number violation was observed; however, even in such a case it would be extremely difficult to distinguish among even the simplest models, let alone more complicated settings).

There are two general reasons behind this unsatisfactory situation:
\begin{enumerate}
\item
The main quantities governing the proton lifetime estimates in GUTs, in particular, the unification scale $M_{\rm GUT}$ (which, in the non-supersymmetric context enters the rates quartically) and the flavour structure of the relevant baryon and lepton number violating currents, are very difficult to estimate with good-enough accuracy from just the low-energy data we have an access to. As for the former, at least a two-loop renormalization-group-equation (RGE) analysis is necessary to keep the error in $M_{\rm GUT}$ at a reasonable level which, however, assumes a detailed knowledge of the relevant threshold corrections (and, hence, the theory spectrum); for the latter one often needs information that is inaccessible even in principle at the electroweak scale, such as, for instance, the shape of the right-handed (RH) rotations in flavour space.
\item Since the unification scale turns out to be only a few orders of magnitude below the Planck scale $M_{\rm Pl}$, the (a-priori unknown) effects of the $M_{\rm Pl}$-suppressed higher-dimensional operators need not be negligible~\cite{Hill:1983xh,Shafi:1983gz,Calmet:2008df}; in practice, they often turn out to be comparable in size to those of the one-loop thresholds and, as such, the associated uncertainties tend to ruin the efforts to go beyond the leading order in precision anyway.
\end{enumerate}

Nevertheless, there seems to exist an exception to these empirical rules, namely, the minimal renormalizable $SO(10)$ grand unified theory~\cite{Chang:1984qr,Deshpande:1992au,Bertolini:2012im} in which the unified gauge symmetry is spontaneously broken by the 45-dimensional scalar. This choice turns out to be rather special as it inhibits the most dangerous class of the leading  order (i.e., $d=5$) gravity-induced operators and, hence, also the corresponding theoretical uncertainties in the determination of $M_{\rm GUT}$.

Remarkably enough, as interesting as it sounds, this scenario has not been considered for more than 30 years since its first formulation at the beginning of 1980's due to notorious tachyonic instabilities~\cite{Yasue:1980fy,Anastaze:1983zk,Babu:1984mz} appearing in its scalar sector along essentially all potentially realistic symmetry breaking chains; it was only in 2010 that these were shown to be just artefacts of the tree-level  approach~\cite{Bertolini:2009es} and that the theory may be fully consistent at the quantum level.

To this end, the simplest version of the relevant Higgs model in which the rank of the gauge group is reduced by a 16-dimensional scalar field has been thoroughly studied in the same work. However, it turns out that the $45\oplus 16$ scenario can hardly support a potentially realistic theory because it is unclear how it could  accommodate the electroweak data (namely, the weak mixing angle) together with (a variant of) the seesaw mechanism for the neutrino masses.
This is namely due to the fact that the seesaw requires two $B-L$ breaking vacuum expectation value (VEV) insertions (recall that the Standard Model singlet in $16$ carries only one unit of $B-L$); this, however, calls for the $B-L$ breaking to occur at a relatively large scale (in the $10^{14}\GeV$ ballpark) which is generically difficult to reconcile with the gauge unification constraints\footnote{This is almost obvious in the minimally fine-tuned scenarios; however, admitting accidentally light extra scalars in the $45\oplus 16$ model does not seem to work either as there are simply no fields around that may affect significantly the running of the strong coupling to the extent achieved, e.g., by the $(8,2,+\tfrac{1}{2}$) scalar in the  $45\oplus 126$ setting.}. For the same reason, the renormalizable alternative due to Witten~\cite{Witten:1979nr} does not work either due to the extra two-loop suppression. Furthermore, it is very problematic to get any firm grip on the flavour structure of this model as any potentially realistic variant of its Yukawa sector relies on a number of contributions from non-renormalizable operators.

Therefore, the most promising scenario of this kind  includes one copy of the $126$-dimensional representation in the scalar sector instead of the spinorial 16; its main virtue is that it can support the standard seesaw mechanism, as well as a potentially realistic (yet simple) Yukawa pattern at the renormalizable level and, thus, avoid most of the aforementioned complications.

The first attempt to study the quantum version of the $45\oplus 126$ model was undertaken in the works \cite{Bertolini:2012im,Bertolini:2013vta,Kolesova:2014mfa} where it was shown that, under several simplifying assumptions, there are extended regions in its parameter space that can support a stable Standard Model (SM) vacuum, accommodate all the SM data and, at the same time, maintain compatibility with the existing proton lifetime constraints. Remarkably, this can all be attained with only a single fine-tuning of the model parameters ensuring one specific heavy scalar in the desert\footnote{Given that the seesaw scale $\sigma$ in all identified potentially realistic settings turns out to be relatively close to $M_{\rm GUT}$ one may even view the situation with the accidentally light scalar $S$ as if the `usual' fine-tuning in $\sigma$ was `traded' for that in its mass $m_{S}$.}~\cite{Bertolini:2013vta,Kolesova:2014mfa}.
The main drawback of these early studies lies in the fact that, out of all relevant quantum corrections emerging at one loop, only the simplest universal type has been taken into account in order to stabilize the scalar sector of the theory with minimum efforts. Hence, those results should be regarded as only approximate and the situation clearly calls for a more complete treatment.

In the current paper we partly fill this gap by calculating in great detail the leading one-loop corrections to the masses of the scalar multiplets transforming as $(8,1,0)$ and $(1,3,0)$ under the $SU(3)_{c}\times SU(2)_{L}\times U(1)_{Y}$ SM gauge group in the $45\oplus 126$ Higgs model. We focus our attention solely on these two fields as they are the principal culprits causing the notorious tree-level tachyonic instabilities mentioned above and, thus, their quantum-level behaviour is of our primary concern. In this sense, a thorough analysis of the relevant  radiative corrections represents the first and minimal step towards any future reliable phenomenological analysis of this scenario.

The work is structured as follows: after a short recapitulation of the tree-level shape of the scalar spectrum of the model in Section~\ref{sect:II} we use the effective potential techniques to calculate the zero-momentum one-loop corrections to the masses of $(8,1,0)$ and $(1,3,0)$ (Section~\ref{sect:III}) and cross-check our results by means of two basic methods: first, by inspecting the relevant formulae in various limits where the spectrum takes specific known shapes and, second, by cross-checking the coefficient of the simplest $SO(10)$-invariant contribution against the direct diagrammatic calculation which, for this term, is relatively easy. Furthermore, we use these results to provide a sample point in the parameter space that is not only free from all the aforementioned pathologies but, at the same time, may even support a potentially realistic GUT scenario; we also add several comments on the methods of implementation of the results in a future numerical analysis. Most of the technicalities are deferred to a set of Appendices. Then we conclude.

\section{The minimal $SO(10)$ Higgs model\label{sect:II}}
In what follows, we shall use the symbols $\phi_{ij}$ and $\Sigma_{ijklm}$ (with all Latin indices running from 1 to 10) for the components of the 45-dimensional adjoint and the 126-dimensional self-dual 5-index antisymmetric tensor irreducible representations of the $SO(10)$ gauge group, respectively. Note that in the real basis of the $SO(10)$ both these structures are fully antisymmetric in all their indices and that $\Sigma$ obeys $\Sigma_{ijklm}=-\tfrac{i}{5!}\varepsilon_{ijklmnopqr}\Sigma_{nopqr}$ provided $\varepsilon_{12345678910}=+1$.  Unlike $\phi$, $\Sigma$ is a complex representation and we shall denote the complex conjugated object by $\Sigma^{*}$. For more on the notation regarding these representations see Appendix~\ref{app:tree-spectrum}.
The decompositions of $\phi$ and  $\Sigma$ into irreducible representations of the SM gauge group are listed in Table~\ref{tab:all-SM-reps}.

\begin{table}[h]
\caption{All types of the SM representations $R$ of scalar fields in the $45\oplus126$ Higgs model. The $\mathbb{R}/\mathbb{C}$ column denotes whether the representation is real or complex (implicitly, for a complex $R$, there is an inequivalent conjugate representation $\overline{R}$), the hash sign $\#$ denotes the multiplicity of $R$ (and consequently the dimension of the corresponding block in the full scalar mass matrix) and the dagger $\dagger$ indicates the presence of a would-be Goldstone mode. There are in general $33$ Goldstones contained in 5 different SM multiplets corresponding to the same number of broken $SO(10)$ generators. The ``size'' column enumerates the real degrees of freedom in the representation $R$ (reflected in the number of equivalent
blocks with identical eigenvalues in the properly reordered full mass matrix). Summing the size $\times$  multiplicity over all blocks  yields  $\sum_{i=a}^{s}{\#_i \times \textrm{size}_i = 45 + 126 + 126 = 297}$ real degrees of freedom in total. There are $19$ different SM representations, out of which $11$ appear only in one copy, $5$ are 2-fold degenerate, $2$ are 3-fold degenerate and $1$ appears even 4 times (the singlet block is $4\times 4$; two real singlets and one complex, which can acquire non-zero vevs $\sqrt{3}\omega_b$, $\sqrt{2}\omega_r$ and $\sqrt{2}\sigma$, respectively). Hence, there are in principle $31$ different eigenvalues; since $5$ of them are Goldstone bosons, one is left with only $26$ non-vanishing (and different) eigenvalues. The $G_{422}$ column denotes the origin of the relevant SM representations within the corresponding representations of the Pati-Salam \PS subgroup of $SO(10)$. (Note however, that none of the representations actually contains a singlet under Pati-Salam, so there can be no physical phase characterized by the Pati-Salam $G_{422}$ gauge symmetry).
\label{tab:all-SM-reps}}
\vskip 0.2cm
\centering
\begin{tabular}{c@{\hskip 0.2cm}lrrrr@{\hskip 1cm}l@{\hskip 0.7cm}l}
\toprule
label$\quad$&\makebox[1.5cm][l]{$R\sim G_{321}$}&\makebox[1cm][r]{$\mathbb{R}/\mathbb{C}$}&\makebox[1cm][r]{$\#^{\phantom{\ast}}$}&&size&$R\;\subseteq G_{422}$&{\hskip -0.35cm}$\subseteq SO(10)$\\
\midrule
$\IND{a}$&$(1,3,0)$&$\mathbb{R}$&$1^{\phantom{\dagger}}$&&$3$&$(1,3,1)$&$\phi$\\\addlinespace[\TABSPACE]
$\IND{b}$&$(8,1,0)$&$\mathbb{R}$&$1^{\phantom{\dagger}}$&&$8$&$(15,1,1)$&$\phi$ \\\addlinespace[\TABSPACE]
$\IND{c}$&$(3,2,-\tfrac{5}{6})$&$\mathbb{C}$&$1^\dagger$&&$12$&$(6,2,2)$&$\phi$ \\\addlinespace[\TABSPACE]
$\IND{d}$&$(1,1,+2)$&$\mathbb{C}$&$1^{\phantom{\dagger}}$&&$2$&$(\overline{10},1,3)$&$\Sigma$ \\\addlinespace[\TABSPACE]
$\IND{e}$&$(1,3,-1)$&$\mathbb{C}$&$1^{\phantom{\dagger}}$&&$6$&$(10,3,1)$&$\Sigma$ \\\addlinespace[\TABSPACE]
$\IND{f}$&$(\bar{3},1,+\tfrac{4}{3})$&$\mathbb{C}$&$1^{\phantom{\dagger}}$&&$6$&$(\overline{10},1,3)$&$\Sigma$ \\\addlinespace[\TABSPACE]
$\IND{g}$&$(3,3,-\tfrac{1}{3})$&$\mathbb{C}$&$1^{\phantom{\dagger}}$&&$18$&$(10,3,1)$&$\Sigma$ \\\addlinespace[\TABSPACE]
$\IND{h}$&$(6,3,+\tfrac{1}{3})$&$\mathbb{C}$&$1^{\phantom{\dagger}}$&&$36$&$(10,3,1)$&$\Sigma$ \\\addlinespace[\TABSPACE]
$\IND{i}$&$(\bar{6},1,-\tfrac{4}{3})$&$\mathbb{C}$&$1^{\phantom{\dagger}}$&&$12$&$(\overline{10},1,3)$&$\Sigma$ \\\addlinespace[\TABSPACE]
$\IND{j}$&$(\bar{6},1,-\tfrac{1}{3})$&$\mathbb{C}$&$1^{\phantom{\dagger}}$&&$12$&$(\overline{10},1,3)$&$\Sigma$ \\\addlinespace[\TABSPACE]
$\IND{k}$&$(\bar{6},1,+\tfrac{2}{3})$&$\mathbb{C}$&$1^{\phantom{\dagger}}$&&$12$&$(\overline{10},1,3)$&$\Sigma$ \\\addlinespace[\TABSPACE]
$\IND{l}$&$(1,2,+\tfrac{1}{2})$&$\mathbb{C}$&$2^{\phantom{\dagger}}$&&$4$&$(15,2,2)$, $(15,2,2)^\ast$&$\Sigma$, $\Sigma^\ast$ \\\addlinespace[\TABSPACE]
$\IND{m}$&$(3,2,+\tfrac{7}{6})$&$\mathbb{C}$&$2^{\phantom{\dagger}}$&&$12$&$(15,2,2)$, $(15,2,2)^\ast$&$\Sigma$, $\Sigma^\ast$ \\\addlinespace[\TABSPACE]
$\IND{n}$&$(8,2,+\tfrac{1}{2})$&$\mathbb{C}$&$2^{\phantom{\dagger}}$&&$32$&$(15,2,2)$, $(15,2,2)^\ast$&$\Sigma$, $\Sigma^\ast$ \\\addlinespace[\TABSPACE]
$\IND{o}$&$(\bar{3},1,+\tfrac{1}{3})$&$\mathbb{C}$&$3^{\phantom{\dagger}}$&&$6$&$(6,1,1)$, $(6,1,1)^\ast$, $(\overline{10},1,3)$&$\Sigma$, $\Sigma^\ast$, $\Sigma$\\\addlinespace[\TABSPACE]
$\IND{p}$&$(1,1,+1)$&$\mathbb{C}$&$2^\dagger$&&$2$&$(1,1,3)$, $(\overline{10},1,3)$&$\phi$, $\Sigma$ \\\addlinespace[\TABSPACE]
$\IND{q}$&$(\bar{3},1,-\tfrac{2}{3})$&$\mathbb{C}$&$2^\dagger$&&$6$&$(15,1,1)$, $(\overline{10},1,3)$&$\phi$, $\Sigma$ \\\addlinespace[\TABSPACE]
$\IND{r}$&$(3,2,+\tfrac{1}{6})$&$\mathbb{C}$&$3^\dagger$&&$12$&$(6,2,2)$, $(15,2,2)$, $(15,2,2)^\ast$&$\phi$, $\Sigma$, $\Sigma^\ast$ \\\addlinespace[\TABSPACE]
$\IND{s}$&$(1,1,0)$&$\mathbb{R}$&$2^{\phantom{\dagger}}$&\rdelim\}{2}{8mm}[$\ 4^\dagger$]&$1$&$(15,1,1)$, $(1,1,3)$,&$\phi$, $\phi$ \\\addlinespace[\TABSPACE]
&&$\mathbb{C}$&$1^{\phantom{\dagger}}$&&$2$&$(\overline{10},1,3)$&$\Sigma$ \\\addlinespace[\TABSPACE]
\bottomrule
\end{tabular}
\end{table}

\subsection{The $SO(10)$ symmetric Lagrangian in the unbroken phase}
The normalization of the component fields in $\phi$ and $\Sigma$  follows the usual convention which fixes the kinetic part of the relevant Lagrangian, $\mathcal{L} = \mathcal{L}_{kin} - V_0$,  to the form
\begin{align}
\mathcal{L}_{kin}& = \tfrac{1}{4} (F_{\mu\nu})_{ij} (F^{\mu\nu})_{ij}  +  \tfrac{1}{4}\,(D_{\mu}\phi)^\dagger_{ij}(D^{\mu}\phi)_{ij}+\tfrac{1}{5!}\,(D_\mu \Sigma)^\dagger_{ijklm}(D^{\mu}\Sigma)_{ijklm},
\end{align}
where
\begin{align}
(F^{\mu\nu})_{ij} & = \partial^{\mu} (A^{\nu})_{ij} - \partial^{\nu} (A^{\mu})_{ij}-i \, g \left[A^{\mu},A^{\nu} \right]_{ij}\,,
\end{align}
and a summation over the repeated Latin indices is implicit.
This yields the ``standard'' kinetic terms for the relevant SM components including coefficients $\tfrac{1}{2}$ and $1$ for real and complex fields, respectively.  
With this at hand, the (renormalizable) scalar potential reads
\begin{align}
V_0(\phi,\Sigma,\Sigma^\ast)&=V_{45}(\phi)+V_{126}(\Sigma,\Sigma^\ast)+V_{\textrm{mix}}(\phi,\Sigma,\Sigma^\ast),\label{eq:scalar-potential}
\end{align}
provided
\begin{align}
V_{45}&=-\frac{\mu^2}{4}(\phi\phi)_0+\frac{a_0}{4}\,(\phi\phi)_0(\phi\phi)_0+\frac{a_2}{4}(\phi\phi)_2(\phi\phi)_2,\\
V_{126}&=-\frac{\nu^2}{5!}(\Sigma\Sigma^\ast)_0+\frac{\lambda_0}{(5!)^2}\,(\Sigma\Sigma^\ast)_0(\Sigma\Sigma^\ast)_0+\frac{\lambda_2}{(4!)^2}(\Sigma\Sigma^\ast)_2 (\Sigma\Sigma^\ast)_2+\nonumber\\
&\quad + \frac{\lambda_4}{(3!)^2(2!)^2}(\Sigma\Sigma^\ast)_4(\Sigma\Sigma^\ast)_4+\frac{\lambda_4'}{(3!)^2}(\Sigma\Sigma^\ast)_{4'}(\Sigma\Sigma^\ast)_{4'}+ \frac{\eta_2}{(4!)^2}(\Sigma\Sigma)_2(\Sigma\Sigma)_2+\frac{\eta_2^\ast}{(4!)^2}(\Sigma^\ast\Sigma^\ast)_2(\Sigma^\ast\Sigma^\ast)_2,\\
V_{\textrm{mix}}&=\frac{i\tau}{4!}(\phi)_2(\Sigma\Sigma^\ast)_2+\frac{\alpha}{2\cdot 5!}(\phi\phi)_0(\Sigma\Sigma^\ast)_0+
\frac{\beta_4}{4\cdot 3!}(\phi\phi)_4 (\Sigma\Sigma^\ast)_4+\frac{\beta_4'}{3!}(\phi\phi)_{4'}(\Sigma\Sigma^\ast)_{4'}+\nonumber\\
&\quad + \frac{\gamma_2}{4!}(\phi\phi)_2 (\Sigma\Sigma)_2+\frac{\gamma_2^\ast}{4!}(\phi\phi)_2 (\Sigma^\ast\Sigma^\ast)_2,\label{eq:full-potential}
\end{align}
where
$
(\phi\phi)_0\equiv\phi_{ij}\phi_{ij}$,
$(\phi\phi)_2 \equiv(\phi\phi)_{jk}\equiv\phi_{ij}\phi_{ik}$,
$(\Sigma\Sigma^\ast)_0\equiv\Sigma_{ijklm}\Sigma^\ast_{ijklm}$,
$(\Sigma\Sigma^\ast)_2\equiv(\Sigma\Sigma^\ast)_{mn}\equiv\Sigma_{ijklm}\Sigma^\ast_{ijkln}$,
$(\Sigma\Sigma^\ast)_4\equiv(\Sigma\Sigma^\ast)_{lmno}\equiv\Sigma_{ijklm}\Sigma^{\ast}_{ijkno}$,
and analogously if $\Sigma$ or $\Sigma^\ast$ is replaced with its conjugate. The invariant contractions among these expressions read
\begin{align}
(\Sigma\Sigma^\ast)_2 (\Sigma\Sigma^\ast)_2&= (\Sigma\Sigma^\ast)_{mn}(\Sigma\Sigma^\ast)_{mn},\nn\\
(\Sigma\Sigma^\ast)_4 (\Sigma\Sigma^\ast)_4&= (\Sigma\Sigma^\ast)_{lmno} (\Sigma\Sigma^\ast)_{lmno},\nn\\
(\Sigma\Sigma^\ast)_{4'} (\Sigma\Sigma^\ast)_{4'}&=(\Sigma\Sigma^\ast)_{lmno} (\Sigma\Sigma^\ast)_{lnmo},\nn\\
(\phi)_2 (\Sigma\Sigma^\ast)_2&=\phi_{mn} (\Sigma\Sigma^\ast)_{mn},\\
(\phi\phi)_4 (\Sigma\Sigma^\ast)_4&=\phi_{lm}\phi_{no}(\Sigma\Sigma^\ast)_{lmno},\nn\\
(\phi\phi)_{4'} (\Sigma\Sigma^\ast)_{4'}&=\phi_{lm}\phi_{no}(\Sigma\Sigma^\ast)_{lnmo},\nn\\
(\phi\phi)_2 (\Sigma\Sigma)_2&=(\phi\phi)_{jk}(\Sigma\Sigma)_{jk}.\nn
\end{align}
Note that there are $3$ parameters with a positive dimension of mass $\{\mu,\nu,\tau\}$ in $V_0$, $9$ dimensionless real parameters $\{a_0,a_2,\lambda_0,\lambda_2,\lambda_4,\lambda_4',\alpha,\beta_4,\beta_4'\}$ and $2$ dimensionless complex parameters $\{\eta_2,\gamma_2\}$. The minus signs in front of $\mu^2$ and $\nu^2$ and the various symmetry factors in other terms are mere convenience. Note also that the coefficient of the $\mu^{2}$ term has been fixed in a different way than in~\cite{Bertolini:2012im,Bertolini:2013vta,Kolesova:2014mfa, Bertolini:2012az}; the slight advantage of the  current notation is the fact that in the symmetric phase $-\mu^{2}$ and $-\nu^{2}$ are exactly the squares of the (tree-level) physical masses of the SM fields in $\phi$ and $\Sigma$, respectively. In what follows, we shall use $\Phi$ as a generic symbol denoting all scalar components at play, i.e., $\Phi\equiv (\phi,\Sigma,\Sigma^{*})$.

\subsection{Spontaneous $SO(10)$ symmetry breaking}
There are $3$ SM singlets in the scalar sector: $2$ real in
$\phi$ and $1$ complex in $\Sigma$. In what follows, we shall denote their potentially non-vanishing VEVs by (for better reading experience concerning, especially, the lengthy formulae in the Appendices we shall use the red color for the VEVs and their simple combinations and blue for the labels of different scalar sector eigenstates, cf. Table~\ref{tab:all-SM-reps}).
\begin{equation}\label{eq:VEVs}
\langle (1,1,1,0)_\phi \rangle  \equiv \sqrt{3}\,\RED{\omega_b},\quad
\langle (1,1,3,0)_\phi \rangle  \equiv \sqrt{2}\,\RED{\omega_r}, \quad
\langle (1,1,3,+2)_\Sigma \rangle =  \langle (1,1,3,-2)_{\Sigma^\ast} \rangle^{*} \equiv \sqrt{2} \RED{\sigma} \;.
\end{equation}
The multiplets above were written in the \LR language and the corresponding fields are assumed to be canonically normalized. The VEVs $\RED{\omega_b}$ and $\RED{\omega_r}$ are real while $\RED{\sigma}$ is a VEV of a complex scalar singlet and, hence, it can be complex. Note that there is a freedom to redefine the overall phase of $\Sigma$ in such a way that $\RED{\sigma}$ can be made real; alternatively, the same transformation can be used to absorb the phase of $\gamma_2$ in equation~\eqref{eq:full-potential}, thus reducing $\gamma_2$ to a real parameter. In the latter case $\sigma$ can be complex (and, hence, it may be convenient to keep track of the relevant complex conjugation as we shall do in what follows).

Assuming no correlations among the VEVs above, the $SO(10)$ gauge symmetry gets spontaneously broken down to the SM group $SU(3)_{c}\times SU(2)_{L}\times U(1)_{Y}$. Special symmetry breaking patterns can be attained in various limits as listed in Table~\ref{tab:groups-in-limits}. From the  phenomenology perspective, however, it is sensible to consider predominantly the case with $|\RED{\sigma}| \ll {\rm max}\{|\RED{\omega_b}|, |\RED{\omega_r}|\}$ in which $\RED{\sigma}$ plays the role of an intermediate (seesaw) scale, while the dominant $\RED{\omega_{r,b}}$ corresponds to the unification scale.
\begin{table}[t]
\caption{Residual gauge symmetries (in a self-explanatory notation) attained for various configurations of the VEVs defined in eq.~\eqref{eq:VEVs}. The last column corresponds to the alternative 'flipped' embedding of the SM hypercharge into the $SU(5)'\times U(1)_{Z'}$ subgroup of the $SO(10)$, cf.~\cite{Barr:1981qv,Derendinger:1983aj}. \label{tab:groups-in-limits}}
\centering
\vskip 0.2cm
\begin{tabular}{llllll}
\toprule
$\phantom{\sigma=\qquad 0}$&$\RED{\omega_b}\neq 0$, $\RED{\omega_r}\neq0\quad\phantom{0}$&$\RED{\omega_b}= 0$, $\RED{\omega_r}\neq0\quad\phantom{0}\quad\phantom{0}$&$\RED{\omega_b}\neq 0$, $\RED{\omega_r}= 0\quad\phantom{0}\quad\phantom{0}$&$\RED{\omega_b}=\RED{\omega_r}\neq 0\quad\phantom{0}$&$\RED{\omega_b}=-\RED{\omega_r}\neq 0\quad\phantom{0}$\\
\addlinespace[\TABSPACE]\midrule
$\RED{\sigma}=0$&$3_c\,2_L\,1_{R}\,1_{B-L}$&$4_c\,2_L\,1_R$&$3_c\,2_L\,2_R\,1_{B-L}$&$5\,1_Z$&$5'\,1_{Z'}$\\\addlinespace[\TABSPACE]
$\RED{\sigma}\neq0$&$3_c\,2_L\,1_Y$&$3_c\,2_L\,1_Y$&$3_c\,2_L\,1_Y$&$5$&$3_c\,2_L\,1_Y$\\\addlinespace[\TABSPACE]
\bottomrule
\end{tabular}
\end{table}

The  tree-level vacuum stability conditions translating among these VEVs and the massive parameters of the potential
read (see also~\cite{Bertolini:2012im})
\begin{align}
\mu^2&=(12 a_0+2a_2)\RED{\omega_b}^2+(8 a_0+2a_2)\RED{\omega_r}^2+2a_2 \RED{\omega_b}\RED{\omega_r}+4 (\alpha +\beta_4')|\RED{\sigma}|^2,\label{eq:mu-tree-level}\\
\nu^2&=3(\alpha+4 \beta_4')\RED{\omega_b}^2+2(\alpha+3 \beta_4')\RED{\omega_r}^2+12 \beta_4' \RED{\omega_b} \RED{\omega_r}+4 \lambda_0 |\RED{\sigma}|^2+a_2 \frac{\RED{\omega_b}\RED{\omega_r}}{|\RED{\sigma}|^2} (\RED{\omega_b}+\RED{\omega_r}) (3\RED{\omega_b}+2\RED{\omega_r}),\\
\tau&=2 \beta_4' (3 \RED{\omega_b}+2 \RED{\omega_r})+a_2 \frac{\RED{\omega_b}\RED{\omega_r}}{|\RED{\sigma}|^2} (\RED{\omega_b}+\RED{\omega_r}).\label{eq:tau-tree-level}
\end{align}
Note that there are potentially problematic terms in the latter two conditions containing $|\RED{\sigma}|^{2}$ in denominators that may\footnote{To this end, let us note that some of the tree-level scalar sector mass-squares calculated  in Appendix~\ref{app:treespectrum} are linear in the combination~(\ref{eq:tau-tree-level}) so the scalar spectrum would be badly distorted if the condition~(\ref{eq:perturbativity}) was not satisfied.} ruin the perturbative expansion whenever the relevant expression exceeds significantly the GUT scale (i.e., the maximum of $\RED{\omega_{b,r}}$). Hence,  in realistic settings one should assume that
\be\label{eq:perturbativity}
a_2 \frac{\RED{\omega_b}\RED{\omega_r}}{|\RED{\sigma}|^2} (\RED{\omega_b}+\RED{\omega_r})\ll M_{\rm Pl}\,.
\ee
\subsection{The tree-level spectrum}\label{sect:tree-level-spectrum}
With this information at hand, the tree-level scalar and gauge spectra of the $45\oplus126$ $SO(10)$ Higgs model under consideration can be readily obtained. Since $45$ is a real representation and $126$ is complex, the total number of real degrees of freedom in the scalar sector is $297$.

For later convenience, it is useful to arrange the second derivatives of $V_{0}$ into a (297-dimensional) Hermitian matrix
\begin{align}
\mathbf{M}_\mathbf{S}^2(\Phi)&\equiv\mathbf{M}_\mathbf{S}^2(\phi,\Sigma,\Sigma^\ast)=\partial\partial^\ast V_{0}(\phi,\Sigma,\Sigma^\ast)=
\begin{pmatrix}
V_{\phi\phi}&V_{\phi\Sigma^\ast}&V_{\phi\Sigma}\\
V_{\Sigma\phi}&V_{\Sigma\Sigma^\ast}&V_{\Sigma\Sigma}\\
V_{\Sigma^\ast\phi}&V_{\Sigma^\ast\Sigma^\ast}&V_{\Sigma^\ast\Sigma}\\
\end{pmatrix}\label{eq:scalar-mass-matrix},
\end{align}
with sub-blocks indicating the types of fields with respect to which the relevant derivatives are taken. In the SM vacuum (characterized by one of the four relevant VEV configurations in the 2nd row of Table~\ref{tab:groups-in-limits}) this matrix encodes the tree-level scalar spectrum of the model and, as such, it may be brought into a block-diagonal form with the non-zero clusters corresponding to the subspaces spanning the irreducible SM representations listed in Table~\ref{tab:all-SM-reps}. Note that the fields of our main interest, i.e., the pseudo-Goldstones $(1,3,0)$ and $(8,1,0)$, are then fully contained in the $V_{\phi\phi}$ sector of $\mathbf{M}_\mathbf{S}^2(\Phi)$.
The complete structure of $\mathbf{M}_\mathbf{S}^2(\Phi)$ in the (block-diagonal) SM basis evaluated at the SM vacuum is given in Appendix~\ref{app:treespectrum}, see also~\cite{Bertolini:2012im}. In order to conform to the needs of the subsequent quantum-level analysis the notation here has been slightly amended\footnote{Besides the overall compactness of the results obtained in Sect.~\ref{sect:III} the new notation facilitates their cross-checking in various limits corresponding to enhanced gauge symmetries, in particular, those listed in Table~\ref{tab:groups-in-limits}.} with respect to that used in~\cite{Bertolini:2012im}; see, in particular,  definitions~\eqref{eq:define-functions-1-begin}--\eqref{eq:define-functions-1-end}.

In a similar manner one can define the 45-dimensional field-dependent mass matrix for gauge bosons $\mathbf{M}_\mathbf{G}^2(\Phi)$, see Appendix~\ref{app:treegaugebosons}. Since we do not consider the breaking of the Standard Model gauge group, this matrix evaluated at the SM vacuum has $12$ massless modes corresponding to the gluons, the $W^{\pm}$, $Z^{0}$ and the photon.

\subsubsection{The masses of the $(1,3,0)$ and $(8,1,0)$ pseudo-Goldstone bosons}
The multiplets of our main interest in the current study are the two scalar pseudo-Goldstone bosons (cf.~\cite{Bertolini:2009es}) transforming as $(1,3,0)$ and $(8,1,0)$ under the $SU(3)_{c}\times SU(2)_{L}\times U(1)_{Y}$ of the SM. Their masses, at the tree level, are (see Appendix~\ref{app:treespectrum} for the notation)
\bea
\label{eq:pgbMass1}
M^{2}_{{\IND{a}}}&=&2\PARtm{a_2}(\RED{\omega_{r}}-\RED{\omega_{b}})(\RED{\omega_{b}}+2 \RED{\omega_{r}})\equiv+2\PARtm{a_2}\RED{\omega'}w^{1}_{[1,2]}\,, \\
\label{eq:pgbMass2}
M^{2}_{{\IND{b}}}&=&2\PARtm{a_2}(\RED{\omega_{b}}-\RED{\omega_{r}})(\RED{\omega_{r}}+2\RED{\omega_{b}})\equiv-2\PARtm{a_2}\RED{\omega'}w^{1}_{[2,1]}\,,
\eea
and thus the scalar spectrum can be non-tachyonic if and only if
\be
\PARtm{a_2} > 0 \quad\text{and}\quad -2<\frac{\RED{\omega_{b}}}{\RED{\omega_{r}}}<-\frac{1}{2}\,,
\ee
i.e., in the vicinity of the flipped $SU(5)$ limit (the 5th column in Table~\ref{tab:groups-in-limits}). This means, however, that the symmetry-breaking chains supporting, at the tree level, a locally convex minimum (i.e., a potentially stable SM-like vacuum) all feature an approximate $SU(5)'\times U(1)_{Z'}$-symmetric scalar spectrum clustering around the superheavy breaking scale $|\RED{\omega_{r}}|\approx |\RED{\omega_{b}}|$, at odds with the gauge unification constraints (at least in the minimally fine-tuned scenarios, i.e., those obeying the minimal survival hypothesis, cf.~\cite{Aulakh:1999pz,Chang:1984qr,Deshpande:1992em,Bertolini:2009qj}).
This, obviously, disqualifies the minimal (and minimally fine-tuned) setting from any potentially realistic model building, at least at the lowest order in  perturbation theory\footnote{To this end, let us reiterate that this conclusion applies even to the situation when the minimal Higgs model is further extended by an extra 10-dimensional scalar multiplet in order to support a viable Yukawa sector with at least two different complex symmetric matrices of Yukawa couplings -- the extra $10_{H}$ does not alter the symmetry breaking pattern by more than just an admixture of its weak doublet component within the electroweak Higgs and an extra set of superheavy doublets and color triplets clustered around the $SO(10)$ breaking scale.}.

Barring, for the sake of this study, the option of non-minimally fine-tuned settings, the only chance to bring the current scenario back from oblivion seems to be a careful inspection of its quantum structure. The hope is that higher order effects may disentangle the overly strong correlation between the two pseudo-Goldstone masses above, at least in case that the tree-level contributions happen to be accidentally small;  this option has been identified (but never inspected in  detail) in previous studies like~\cite{Bertolini:2009es}. A detailed calculation of the relevant radiative contributions to the  tree-level mass relations ~\eqref{eq:pgbMass1} and~\eqref{eq:pgbMass2} is the scope of the next section.

\section{One-loop pseudo-Goldstone masses in the minimal $SO(10)$ Higgs model\label{sect:III}}
\subsection{One-loop scalar masses from the effective potential}

In this section we review some of the technical aspects of the effective potential formalism we adopt for the computation of the desired scalar masses at the 1-loop level.

\subsubsection{Scalar mass matrix at the one-loop level}\label{sect:CW-mechanism}
In the effective potential approach there are in general two types of effects  contributing to the scalar masses at the one-loop level (i.e., at the order characterized by one power of the generic $\hbar/{16 \pi^2}$ suppression factor), namely:
\begin{enumerate}
\item
The usual one-loop corrections to the two-point 1PI Green's functions whose roots in the true vacuum define the pole masses of the scalar excitations.
\item The quantum shift of the vacuum which justifies the use of a simplified perturbation theory in which there are no degenerate one-point vertices in the interaction part of the Lagrangian density (aka ``tadpole cancellation'').
\end{enumerate}
Up to the first power in $\hbar$, the relevant combination of the two effects governing the $\hbar$-expansion of the one-loop scalar mass matrix $m$ (in the zero-momentum scheme implicitly assumed within the effective potential approach) reads formally
\begin{align}
m^2_{ab}&\equiv\partial_a\partial_b V|_{v}=\partial_a\partial_b V_0|_{v_0+\hbar \, v_1}+\hbar \, \partial_a \partial_b V_1|_{v_0}+\mathcal{O}(\hbar^2)\,,\label{eq:1loop-masses-decomposition}
\end{align}
where
$
V = \sum_{i=0}^{\infty} V_i \, \hbar^{i} = V_0 + \hbar \, V_1 + \mathcal{O}(\hbar^2)
$ is the Coleman-Weinberg effective potential~\cite{Coleman:1973jx} and $v = \sum_{i=0}^{\infty} v_i \, \hbar^{i} = v_0+\hbar \, v_1+\mathcal{O}(\hbar^2)$ denotes the true quantum vacuum of the theory determined from the stationary point condition
\begin{align}
\partial_a V|_{v}&= \partial_a V_0|_{v} + \hbar \, \partial_a V_1|_{v}= 0\label{eq:1loop-vacuum-condition}\,.
\end{align}
Beside the tree-level contribution $V_{0}$ discussed at length in the previous section, c.f.~\eqref{eq:scalar-potential}, the one-loop scalar potential (in the zero-momentum scheme) is given by
\begin{align}
V_1(\Phi,\mu_r)&=\frac{1}{64\pi^2}\,\Tr\left[\mathbf{M}_\mathbf{S}^4(\Phi)\left(\log\frac{\mathbf{M}_\mathbf{S}^2(\Phi)}{\mu_r^2}-\frac{3}{2}\right)+3\mathbf{M}_\mathbf{G}^4(\Phi)\left(\log
\frac{\mathbf{M}_\mathbf{G}^2(\Phi)}{\mu_r^2}-\frac{5}{6}\right)\right],\label{eq:Coleman-Weinberg}
\end{align}
where
$\mathbf{M}_\mathbf{S}^2(\Phi)$ and $\mathbf{M}_\mathbf{G}^2(\Phi)$ are the tree-level field-dependent scalar and gauge mass matrices introduced in Sect.~\ref{sect:tree-level-spectrum} (with boldface always denoting matrix structures) and $\mu_r$ is the relevant renormalization scale.

The quantum-level contribution to the stationary point condition~\eqref{eq:1loop-vacuum-condition} is then given by
\begin{align}
\partial_a V_1&= -\frac{1}{32\pi^2}\;\Tr\left[\mathbf{M}_\mathbf{S}^2 \,\partial_a \mathbf{M}_\mathbf{S}^2+\mathbf{M}_\mathbf{G}^2 \,\partial_a \mathbf{M}_\mathbf{G}^2\right]
 +\frac{1}{64\pi^2}\;\Tr\left[\left\{\mathbf{M}_\mathbf{S}^2,\partial_a \mathbf{M}_\mathbf{S}^2\right\}\log\frac{\mathbf{M}_\mathbf{S}^2}{\mu_r^2}+3\left\{\mathbf{M}_\mathbf{G}^2,\partial_a \mathbf{M}_\mathbf{G}^2 \right\}\log\frac{\mathbf{M}_\mathbf{G}^2}{\mu_r^2}\right]\,;\label{eq:1loop-vacuum}
\end{align}
note that we have dropped all the brackets denoting the implicit dependence of the mass matrices on the scalar fields of the model.

Due to the general non-commutativity of ${\bf M}_{\bf S,G}^{2}$ with their own first derivatives the second derivative of the formula~\eqref{eq:Coleman-Weinberg} is far more involved:
\begin{align}
\partial_a \partial_b V_1&=-\frac{1}{32\pi^2}\Tr\left[\partial_a \mathbf{M}_\mathbf{S}^2\,\partial_b \mathbf{M}_\mathbf{S}^2+\mathbf{M}_\mathbf{S}^2 \,\partial_a\partial_b \mathbf{M}_\mathbf{S}^2+\partial_a \mathbf{M}_\mathbf{G}^2\,\partial_b \mathbf{M}_\mathbf{G}^2+\mathbf{M}_\mathbf{G}^2\,\partial_a\partial_b \mathbf{M}_\mathbf{G}^2\right]\nonumber\\
&\quad +\frac{1}{64\pi^2}\,\Tr\left[\left(\left\{\partial_a \mathbf{M}_\mathbf{S}^2,\partial_b \mathbf{M}_\mathbf{S}^2\right\}+\left\{\mathbf{M}_\mathbf{S}^2,\partial_a\partial_b \mathbf{M}_\mathbf{S}^2\right\}\right)\,\log\frac{\mathbf{M}_\mathbf{S}^2}{\mu_r^2}+\mathbf{S}_{ab}\right]\nonumber\\
&\quad +\frac{3}{64\pi^2}\,\Tr\left[\left(\left\{\partial_a \mathbf{M}_\mathbf{G}^2,\partial_b \mathbf{M}_\mathbf{G}^2\right\}+\left\{\mathbf{M}_\mathbf{G}^2,\partial_a\partial_b \mathbf{M}_\mathbf{G}^2\right\}\right)\,\log\frac{\mathbf{M}_\mathbf{G}^2}{\mu_r^2}+\mathbf{G}_{ab}\right]\,.
\label{eq:1loop-masses}
\end{align}
Here
\begin{align}
\mathbf{S}_{ab}=\mathbf{\Upsilon}\left(\frac{\mathbf{M}_\mathbf{S}^2}{\mu_r^2},\partial_a \mathbf{M}_\mathbf{S}^2,\partial_b \mathbf{M}_\mathbf{S}^2\right ),\qquad
\mathbf{G}_{ab}=\mathbf{\Upsilon}\left(\frac{\mathbf{M}_\mathbf{G}^2}{\mu_r^2},\partial_a \mathbf{M}_\mathbf{G}^2,\partial_b \mathbf{M}_\mathbf{G}^2\right ),
\label{eq:SG}
\end{align}
are expressed via a matrix function $\mathbf{\Upsilon}$ including an infinite series of nested commutators
\begin{align}
\mathbf{\Upsilon}(\mathbf{A},\mathbf{A}_a,\mathbf{A}_b)& =\sum_{m=1}^{\infty}(-1)^{m+1}\,\frac{1}{m}\sum_{k=1}^{m}\binom{m}{k}\left\{\mathbf{A},\mathbf{A}_a\right\}\underbrace{\left[\mathbf{A},\ldots
\left[\mathbf{A},\mathbf{A}_b\right]\ldots\right]}_{(k-1)\times\,\textrm{commutator}}(\mathbf{A}-\mathbb{1})^{m-k} \label{eq:nested-commutator-series},
\end{align}
where the first commutator bracket is just $\mathbf{A}_b$, the second is $\left[\mathbf{A},\mathbf{A}_b\right]$, the third $\left[\mathbf{A},\left[\mathbf{A},\mathbf{A}_b\right]\right]$ and so on.  The general strategy for dealing with the formula~\eqref{eq:nested-commutator-series}, together with a brief discussion of the shape of the results it yields, is given in Section~\ref{sect:methods}.

Let us also remark that there are no such issues in the expression~\eqref{eq:1loop-vacuum} because of the cyclic property of the overall trace which admits a resummation of the ``raw'' expression into the simplified matrix logarithm representation.

\subsubsection{Dealing with the nested commutators \label{sect:methods}}
In this section we describe several tricks that facilitate dealing with the nested commutators\footnote{This concerns namely the $\mathbf{S}_{ab}$ structure defined in eq.~\eqref{eq:SG} coming from the huge scalar sector; the gauge contribution proportional to $\Tr\mathbf{G}_{ab}$ turns out to be very simple since the $45\times 45$ gauge boson mass matrix evaluated in the vacuum is almost diagonal, see Table~\ref{tab:gauge-boson-tree-masses}.}, focusing mainly on their numerical evaluation; as it turns out, a full analytic account is tractable only in special cases as, for instance, the one discussed in Section~\ref{sect:results}. For all the derivations and proofs of the expressions in use see Appendix~\ref{app:nested-commutator-evaluation}.

The object of our main interest, i.e., the trace of $\mathbf{\Upsilon}$ evaluated in the tree-level vacuum, cf. \eqref{eq:1loop-masses}--\eqref{eq:SG}, may be further simplified using the identity (see Appendix~\ref{app:nested-commutator-evaluation})
\begin{align}
\Tr\mathbf{\Upsilon}(\mathbf{A},\mathbf{A}_a,\mathbf{A}_b)&=\sum_{i,j;\;\lambda_i\neq \lambda_j}\!\!M^{a}_{ji}M^{b}_{ij}\,\frac{\lambda_i+\lambda_j}{\lambda_i-\lambda_j}\,\log\frac{\lambda_i}{\lambda_j} + \sum_{i,j;\;\lambda_i=\lambda_j}\!\!2M^{a}_{ji}M^{b}_{ij},\label{eq:general-nested-commutator}
\end{align}
where $\lambda_i$ are eigenvalues of the matrix $\mathbf{A}$ with corresponding orthonormal eigenvectors $\mathbf{v}_i$, while $M^a$ and $M^{b}$ are the matrices $\mathbf{A}_a$ and $\mathbf{A}_b$ rotated into the orthonormal eigenbasis of $\mathbf{A}$:
\begin{align}
M^{a}_{ij}&=\mathbf{v}_i^\dagger\mathbf{A}_{a}\mathbf{v}_j,\label{eq:definition-Ma}\\
M^{b}_{ij}&=\mathbf{v}_i^\dagger\mathbf{A}_{b}\mathbf{v}_j.\label{eq:definition-Mb}
\end{align}
Let us note that this approach is fully general and applicable (at least numerically) for any form of the $\mathbf{A}$, $\mathbf{A}_{a}$ and $\mathbf{A}_{b}$ matrices. In that sense, it is superior to the method used previously in, e.g., ref.~\cite{Malinsky:2012tp} which assumed a simple geometric behaviour of the nested commutators from a certain value of the $k$ index in eq.~\eqref{eq:nested-commutator-series} onwards -- unfortunately, unlike in the case of the simpler $45\oplus 16$ model studied previously in~\cite{Bertolini:2009es} (or the Abelian Higgs model, cf.~\cite{Malinsky:2012tp}) where this was indeed the case, the situation in the $45\oplus 126$ Higgs model is more complicated.

Another point worth making here concerns the visual difference between the two contributions in eq.~\eqref{eq:general-nested-commutator} -- the former structure, belonging to a set of non-degenerate eigenvalues, suggests a log-type behaviour while the latter tends to yield non-log terms (in fact, in most cases even polynomials). Since the same two types of terms emerge also from the ``non-commutator'' parts of the basic formula~\eqref{eq:1loop-masses}, the distinction between log and non-log terms is in fact very handy and we shall use it in the next section.
However, it becomes rather fuzzy when it comes to limits in which  the character of the spectrum changes qualitatively, i.e., when the degeneracies increase. Indeed, if two formerly non-degenerate eigenvalues $\lambda_{i,j}$ become equal in a certain limit, one has
\begin{align}
\lim_{\lambda_i\to\lambda_j}\,(\lambda_i+\lambda_j)\,\frac{\log\lambda_i-\log\lambda_j}{\lambda_i-\lambda_j}&=2\,,
\label{eq:metamorphosis}
\end{align}
and hence a term of the first type becomes formally a second-type contribution, cf. Section~\ref{sec:checks-limits}.

\subsubsection{One-loop masses of the $(1,3,0)$ and $(8,1,0)$ scalars in the $45\oplus 126$ Higgs model}\label{sect:results}
The one-loop (zero-momentum-scheme) masses of the $(1,3,0)$ and $(8,1,0)$ scalars can be written as
\begin{align}
M^{2}_{\IND{a},\text{1-loop}}&=M^{2}_{\IND{a}}+\Delta_{\IND{a}}^{G[\textrm{poly}]}+\Delta_{\IND{a}}^{G[\textrm{log}]}+\Delta_{\IND{a}}^{S_{FIN}[\textrm{poly}]}+\Delta_{\IND{a}}^{S_{INF}[\textrm{poly}]}+\Delta_{\IND{a}}^{S[\textrm{log}]}\,,\label{eq:mass1-all-triplet}\\
M^{2}_{\IND{b},\text{1-loop}}&=M^{2}_{{\IND{b}}}+\Delta_{\IND{b}}^{G[\textrm{poly}]}+\Delta_{\IND{b}}^{G[\textrm{log}]}+\Delta_{\IND{b}}^{S_{FIN}[\textrm{poly}]}+\Delta_{\IND{b}}^{S_{INF}[\textrm{poly}]}+\Delta_{\IND{b}}^{S[\textrm{log}]}\,, \label{eq:mass1-all-octet}
\end{align}
\noindent
where the $\Delta$ symbols correspond to different types of the one-loop contributions calculated from the formulae~\eqref{eq:1loop-masses}--\eqref{eq:nested-commutator-series}.  These are clustered with respect to their origin (gauge with superscript $G$ and scalar with superscript $S$) and their mathematical form (polynomial [poly] and logarithmic [log]) as follows:
\begin{enumerate}
\item The $\Delta^{G[\textrm{poly}]}$ contributions contain all the polynomial-type terms from the graphs with gauge bosons running in the loops. As such, these contributions are all proportional to $g^4$, where $g$
is the $SO(10)$ gauge coupling constant. They generally come from both terms on the RH side of eq.~\eqref{eq:1loop-masses-decomposition}, including the polynomial part of the $1$-loop vacuum substituted into $\partial^{2}V_0$, the polynomial terms of $\partial^{2}V_1$ given in eq.~\eqref{eq:1loop-masses} and also the second type of terms in the expression for the nested commutator series~\eqref{eq:general-nested-commutator} corresponding to degenerate eigenvalues of $\mathbf{M}_\mathbf{G}^2$.
\item The $\Delta^{G[\textrm{log}]}$ terms contain all the logarithmic terms from the diagrams with gauge bosons running in the loops. As before, there are $g^{4}$ proportionality factors in front of the logs whose arguments
are the masses of the massive gauge bosons corresponding to the  broken generators of $SO(10)$. Again, these come from both parts of expression~\eqref{eq:1loop-masses-decomposition} including, in this case, the ``non-degenerate'' contributions from eq.~\eqref{eq:general-nested-commutator}.
\item The $\Delta^{S_{FIN}[\textrm{poly}]}$ terms contain all polynomial contributions from the scalars running in the loop, except for those coming from the nested commutator series (i.e., they are fully contained in the  $\partial^{2}V_0$ factor in~\eqref{eq:1loop-masses-decomposition} and the ``finite'' part of the formula~\eqref{eq:1loop-masses}.
These terms are homogenous quadratic polynomials in the parameters of the scalar potential~\eqref{eq:scalar-potential}.
\item The $\Delta^{S_{INF}[\textrm{poly}]}$ pieces denote the polynomial scalar contributions coming solely from the infinite series terms $\mathbf{S}_{ab}$ in eq.~\eqref{eq:1loop-masses}  where they emerge from the scalar spectrum degeneracies due to the residual  Standard Model gauge symmetry.
\item Finally, the $\Delta^{S[\textrm{log}]}$ structure labels all the logarithmic contributions associated to the graphs with scalars running in the loop. These again come from both parts of the expression~\eqref{eq:1loop-masses-decomposition}, including the ``non-degenerate'' contributions from eq.~\eqref{eq:general-nested-commutator}. The coefficients in front of the logs are homogenous quadratic polynomials of the tree-level scalar potential parameters, while their arguments contain the (squared) tree-level masses of the relevant scalars in the loops.
\end{enumerate}
Note that the $\Delta^{S_{INF}[\textrm{poly}]}$ terms have been singled out because these are rather difficult to calculate analytically and the results in a closed form are cumbersome; the same applies also to $\Delta^{S[\textrm{log}]}$. For this reason, we shall not present them in their full complexity but rather in a simplified form that they attain in the limit\footnote{Note that in the $\sigma \to 0$, $\gamma_2 \to 0$ limit the $\Sigma$-self interaction terms do not contribute to the1-loop masses of the fields coming solely from $\phi$. Hence, one can neglect the $\lambda_0$, $\lambda_2$, $\lambda_4$, $\lambda_4'$ and $\eta_2$ terms in the potential $V_0$ from the beginning (in fact, $\eta_2$ is absent from $M^2_a$ and $M^2_b$ even at the level of field dependent tree-level mass matrices). The reason is that all the off-diagonal blocks in the mass matrix~\eqref{eq:scalar-mass-matrix} in vacuum vanish in this limit. Then the only mixing within the 126-dimensional $V_{\Sigma\Sigma^\ast}$ block is among the states in $M_{\IND{o}}^2$ belonging either to $(6,1,1)$ or $(\overline{10},1,3)$ of \PS.}
\begin{align}
\RED{\sigma}\to 0,\quad a_{2}&\to 0,\quad \gamma_{2}\to 0,\quad \frac{a_2}{|\RED{\sigma}|^{2}}=\textrm{const}.\label{eq:G3211-limit-original}
\end{align}
Let us also remark that this setting is not an arbitrary choice but rather a physically well-motivated approximation to the general case: as far as the $\sigma\to 0$ limit is concerned, the ``delayed breaking'' of the $U(1)_{B-L}$ obtained in such situation corresponds to the potentially realistic seesaw scale with the RH neutrino masses well below $M_{\rm GUT}$; the $a_{2}\to 0$ and $\gamma_{2}\to 0$ limits are, on the contrary, suggested by the (simplified) preceding studies~\cite{Bertolini:2013vta,Kolesova:2014mfa} as the only situation in which a fully non-tachyonic spectrum compatible with the gauge unification constraints seems to be attainable.

The full analytic form (modulo the aforementioned limit~\eqref{eq:G3211-limit-original} adopted for simplicity reasons for the $\Delta^{S_{INF}[\textrm{poly}]}$ and $\Delta^{S[\textrm{log}]}$ pieces) of the one-loop corrections entering formulae~\eqref{eq:mass1-all-triplet}--\eqref{eq:mass1-all-octet} is given in Appendix~\ref{app:full-one-loop-contributions}.
\subsubsection{Going to the mass shell}\label{sect:mass-shell}
As we have already mentioned, the formulae~\eqref{eq:mass1-all-triplet}--\eqref{eq:mass1-all-octet} with the $\Delta$ factors given in Appendix~\ref{app:full-one-loop-contributions} encode the masses of the two pseudo-Goldstone bosons of our interest in the zero-momentum renormalization scheme. This, however, is potentially problematic for at least two reasons:
\begin{enumerate}
\item There are peculiar infra-red (IR) divergences due to a certain number of zero eigenvalues in the arguments of logs in the $\Delta^{S,G[\textrm{log}]}$ terms above when the tree-level field-dependent mass matrix ${\bf M}_{\bf S}^{2}(\Phi)$ in~\eqref{eq:1loop-masses} is evaluated at the tree-level vacuum\footnote{Remarkably, all the spurious IR divergences happen to disappear from the triplet and octet $\Delta$ factors in the limit \eqref{eq:G3211-limit-original}.}. Obviously, these are associated with the Goldstone modes whose propagators, in the Landau gauge, have poles at $p^{2}=0$.
\item On the practical side, these masses should be eventually used as inputs of a dedicated phenomenological analysis including constraints from two-loop gauge unification requirements. The relevant calculations are, however, most conveniently performed in different schemes such as the $\overline{\rm MS}$ and, hence, their inputs should be adopted to the same scheme for consistency reasons.
\end{enumerate}
A minimal and natural solution to both these issues is provided by   the transition from the zero-momentum to the on-shell scheme in which the physical masses are given as a solution of the secular equation
\be\label{eq:massshell}
\det\left[p^{2}-m^{2}-\Sigma(p^{2})+\Sigma(0)\right]=0\,,
\ee
where $m^{2}$ is the matrix of the second derivatives of the effective potential in the vacuum calculated above and $\Sigma(p^{2})$ denotes the corresponding matrix of the scalar fields' self-energies (in any scheme; the scheme dependence of $\Sigma$ drops out of the difference above). Note also that, by definition, $\Sigma(0)$ is nothing but the loop part of $m^{2}$.

In principle, the transition from the zero-momentum to the on-shell masses is highly non-trivial as it includes the full structure of $\Sigma(p^{2})$. However, given the scope of this study, i.e., to provide a robust description of the heavy spectrum for a future two-loop RG analysis\footnote{This, in the usual situation, requires just the tree-level masses inserted into the relevant one-loop matching formulae~\cite{Weinberg:1980wa,Hall:1980kf}; however, in models which possess a (meta)stable vacuum supporting a non-tachyonic spectrum only at the loop level, the critical (i.e., potentially tachyonic) sectors of the spectrum require regularization by means of radiative corrections.}, the effects of $m^{2}+\Sigma(p^{2})-\Sigma(0)$ in the calculation of the masses of the fields of our main interest, i.e., the tree-level-tachyonic pseudo-Goldstone bosons, may be still reasonably  approximated by the contributions from $m^{2}$ only if their pole masses stay somewhat below those of the heavy fields ($M$) circulating in the relevant loops. This may be readily seen from the momentum expansion of the typical scalar-field contribution to $\Sigma(p^{2})-\Sigma(0)$:
\be
\Sigma(p^{2})-\Sigma(0)= \frac{1}{16\pi^{2}}\left(c_{1}p^{2}+c_{2}\frac{p^{4}}{M^{2}}+\ldots\right)\,,
\ee
where $c_{i}$ are numerical ${\cal O}(1)$ coefficients with $i$ denoting the power of $p^{2}$ in the numerators of the corresponding terms. Substituting this into~\eqref{eq:massshell} and solving for $p^{2}$ in the regime in which the tree-level contribution to $m^{2}$ is absent or strongly suppressed (with respect to the dominant 1-loop contribution of the order of ${M^{2}}/{16\pi^{2}}$) the on-shell mass, i.e., the physical root of~\eqref{eq:massshell}, obeys $m_{\rm phys}^{2}={\cal O}({M^{2}}/{16\pi^{2}})$. Hence,  $\Sigma(p^{2})-\Sigma(0)= {\cal O}[M^{2}/{(16\pi^{2})^{2}}]={\cal O}(m_{\rm phys}^{4}/M^{2})$
which is clearly subleading with respect to  the leading contribution from $m^{2}$.

The only exception to this simple reasoning is the case when some of the sub-blocks of the tree-level scalar mass matrix in the arguments of the log terms contain Goldstone-mode zeros which are not regulated by the corresponding zero pre-factors. Such an IR divergence is then compensated only by the $\Sigma(0)$ term in eq.~\eqref{eq:massshell} which, however, is much easier to calculate than the full-fledged $\Sigma(p^{2})$; alternatively, one can just discard such IR divergences at the leading order in the perturbative expansion.

In summary, for those fields whose masses are dominated by the one-loop corrections, there is no need to deal with the self-energy at the leading order and the IR-regulated zero-momentum mass expressions derived from the effective potential are sufficient as inputs of a two-loop RGE analysis.

\subsection{Consistency checks}
Given the high complexity of the results presented in Appendix~\ref{app:full-one-loop-contributions} we find it convenient to supply a set of their consistency checks concerning their behaviour in several limits corresponding to an enhanced gauge symmetry when the character of the spectrum changes qualitatively.

\subsubsection{Limits\label{sec:checks-limits}}
There are two specific limits in which one can anticipate the form of the one-loop results~\eqref{eq:mass1-all-triplet} and~\eqref{eq:mass1-all-octet} on the symmetry basis corresponding to the standard and the flipped $SU(5)\times U(1)$ scenarios, respectively, attained in the regimes $\RED{\omega_b}=\pm\RED{\omega_r}$, see Table~\ref{tab:groups-in-limits}.
\subparagraph{The ``standard'' $SU(5)\times U(1)$ limit $\omega_r\to\omega_b$:}
In this limit, the one-loop triplet and octet masses~\eqref{eq:mass1-all-triplet} and~\eqref{eq:mass1-all-octet} should vanish as they do at the tree level, see~\eqref{eq:pgbMass1} and~\eqref{eq:pgbMass2}; the reason is that they become members of an $SU(5)\times U(1)$ multiplet which, in the SM vacuum, contains a Goldstone mode $(3,2,-\tfrac{5}{6})+h.c.$, cf. Sect.~\ref{sect:exactG}.

In order to see this, it is convenient to substitute
$\RED{\omega_r}=\RED{\omega_b}+\kappa$ into the relevant formulae in Appendix~\ref{app:full-one-loop-contributions} and then take the $\kappa\to 0$ limit. Note that the contributions of the logarithmic and polynomial type in~\eqref{eq:mass1-all-triplet} and~\eqref{eq:mass1-all-octet} do not need to vanish separately due to the aforementioned metamorphosis~\eqref{eq:metamorphosis} of some of the logarithmic terms into polynomial form.
The behaviour of the individual contributions to $m^{2}_{{\IND{a,b}},\text{1-loop}}$ is sketched in the following scheme:
\begin{align}
m_{{\IND{a}},\textrm{tree}}^{2},\ m_{{\IND{b}},\textrm{tree}}^{2}&\xrightarrow{\kappa\to 0}0,\nn\\
\Delta_{\IND{a}}^{S_{FIN}[\textrm{poly}]},\  \Delta_{\IND{b}}^{S_{FIN}[\textrm{poly}]}&\xrightarrow{\kappa\to 0}0,\nn\\
\Delta_{\IND{a}}^{G[\textrm{poly}]}+\Delta_{\IND{a}}^{G[\textrm{log}]},\  \Delta_{\IND{b}}^{G[\textrm{poly}]}+\Delta_{\IND{b}}^{G[\textrm{log}]}&\xrightarrow{\kappa\to 0}0,\label{eq:SU5limits}\\
\Delta_{\IND{a}}^{S_{INF}[\textrm{poly}]}+\Delta_{\IND{a}}^{S[\textrm{log}]},\  \Delta_{\IND{b}}^{S_{INF}[\textrm{poly}]}+\Delta_{\IND{b}}^{S[\textrm{log}]}&\xrightarrow{\kappa\to 0}0.\nn
\end{align}
The proof of these equalities is slightly complicated by the fact that all the pre-factors of the log terms tend to blow in the $\kappa\to 0$ limit, see Appendices~\ref{app:gaugeloops} and~\ref{app:scalarloops}, which renders the individual log-type contributions divergent. The obvious trick is to group the logs whose arguments converge to the same limit and use the identity
\begin{align}
\lim_{\kappa\to 0}\left[\sum_{x} \frac{A_x(\kappa)}{\kappa}\,\log\left(m_0+c_x\kappa+\mathcal{O}(\kappa^{2})\right)\right]&=\lim_{\kappa\to 0}\left[\frac{\sum_{x}A_x(\kappa)}{\kappa}\right]\log\left(m_0\right)+\sum_{x}\frac{c_x\,A_x(0)}{m_0},\label{eq:compute-in-limit}
\end{align}
where $x$ sums over a group of indices with the same argument $m_0$ (for $\kappa\to 0$) in the logarithm which, for instance, for the scalar contributions\footnote{Note that there are 22 different log terms in the scalar sector while only 4 of them are generated by the gauge interactions, cf. Appendices~\ref{app:gaugeloops} and~\ref{app:scalarloops}.} are
\begin{align}
\{x\}:&\quad\{\IND{l_2}, \IND{o_1}\}; \{\IND{p}, \IND{q}, \IND{r_1}\};\{\IND{e},\IND{k}, \IND{r_2}\};\{\IND{d}, \IND{h},\IND{i},\IND{m_1},\IND{n_2},\IND{o_3}\};
\{\IND{f}, \IND{g},\IND{j},\IND{l_1}, \IND{m_2}, \IND{n_1},\IND{o_2}\},\label{eq:index-grouping}
\end{align}
 $A_x(\kappa)$ are analytic functions of $\kappa$ and $c_x$ is the first expansion coefficient in $\kappa$ of the logarithm arguments. Only terms with a non-zero constant part in $A_x(\kappa)$ give non-vanishing polynomial pieces in the $\kappa\to 0$ limit, i.e.~only the logs with divergent prefactors can give rise to a polynomial contribution.
The behaviour~\eqref{eq:SU5limits} can be viewed as a rather non-trivial consistency check of the results because the constant and linear terms in $\kappa$ in the sums $\sum_x A_x(\kappa)$ must in all cases drop, as they indeed do.

\subparagraph{The ``flipped'' $SU(5)\times U(1)$ limit $\omega_r\to -\omega_b$:}
In this case, the symmetry group is $\mathrm{SU}(5)'\times\mathrm{U}(1)'$, cf. Table~\ref{tab:groups-in-limits}, the hypercharge is a linear combination of one of the Cartans of $\mathrm{SU}(5)'$ and the extra $\mathrm{U}(1)'$ charge. The triplet $(1,3,0)$ and the octet $(8,1,0)$ again become part of the same 24-dimensional gauge multiplet (this time together with representations $(3,2,+\tfrac{1}{6})+h.c.$). This is easily seen from the form of the $M^{2}_{\IND{r}}$ matrix~\eqref{eq:Mrmatrix} which becomes diagonal in this limit, with the $[11]$ entry therein  reducing  to the same expression as $M^{2}_{\IND{a}}$ and  $M^{2}_{\IND{b}}$ (in the same limit). In contrast with the ``standard'' $\mathrm{SU}(5)$ case above, the $(3,2,+\tfrac{1}{6})+h.c.$ multiplet is not a would-be Goldstone boson (in the SM vacuum), so the octet and triplet masses should be equal but non-vanishing.

As before, it is convenient to implement this limit by means of the $\kappa$ parameter ($\RED{\omega_r}=-\RED{\omega_b}+\kappa$) with $\kappa\to 0$ and, for the scalar contributions, it is  useful to group the logarithmic terms according to the scheme\footnote{These groupings are listed in the same ordering as the log terms in equation~\eqref{eq:scalar-flipped-su5}, with all IR divergences coming from the last grouping of indices. As we have argued in Sect.~\ref{sect:mass-shell} these divergences disappear in the on-shell formula~\eqref{eq:massshell}.}
\begin{align}
\{x\}:&\quad \{\IND{s}\};\{\IND{d}\}; \{\IND{l_1}, \IND{o_1}\}; \{\IND{f},\IND{m_1},\IND{p}\}; \{\IND{e},\IND{i},\IND{m_2}\}; \{\IND{g}, \IND{j}, \IND{l_2}, \IND{n_2}, \IND{o_2}, \IND{q}, \IND{r_2}\};\{\IND{h},\IND{k},\IND{n_1}, \IND{o_3},\IND{r_1}\}.
\end{align}
The results assume the following form:
\begin{align}
m^{2}_{\IND{a},\textrm{tree}},m^{2}_{\IND{b},\textrm{tree}}&\xrightarrow{5'\,1_{Z'}} 4 a_2 \RED{\omega_b}^2, \\
\Delta_{\IND{a}}^{G[\textrm{poly}]},\Delta_{\IND{b}}^{G[\textrm{poly}]}&\xrightarrow{5'\,1_{Z'}} \tfrac{17}{32 \pi ^2}\, g^4 \RED{\omega_b}^2, \\
\Delta_{\IND{a}}^{G[\textrm{log}]},\Delta_{\IND{b}}^{G[\textrm{log}]}&\xrightarrow{5'\,1_{Z'}} \tfrac{3}{32 \pi ^2}\, g^4 \RED{\omega_b}^2 \log \left[g^2 \RED{\omega_b}^2/\mu_r^2\right], \\
\Delta_{\IND{b}}^{S_{FIN}[\textrm{poly}]},\Delta_{\IND{b}}^{S_{FIN}[\textrm{poly}]}&\xrightarrow{5'\,1_{Z'}} -\tfrac{1}{8\pi^2} \left(96 a_0  a_2 +76 a_2 ^2+560 |\gamma_2|^{2}-5 \beta_4 ^2+60 \beta_4  \beta_4' -100 \beta'_4{}^2 \right)\,\RED{\omega_b}^2.
\\
\begin{array}{cc}
\Delta_{\IND{a}}^{S_{INF}[\textrm{poly}]}+\Delta_{\IND{a}}^{S[\textrm{log}]},
\\
\Delta_{\IND{b}}^{S_{INF}[\textrm{poly}]}+\Delta_{\IND{b}}^{S[\textrm{log}]}
\end{array}
&\xrightarrow{5'\,1_{Z'}} \tfrac{5}{8\pi ^2}\,F\big(7\Separator 0\Separatorb 0\Separator 9 \RED{\omega_b}^2\Separatorb 12 \RED{\omega_b}^2\Separatorb 60 \RED{\omega_b}\big)+\tfrac{1}{16\pi^2}\times \nonumber\\
&\qquad\quad \times\Big(F\big(\!-\!4\Separator 0\Separatorb 64 \RED{\omega_b}\Separator 0\Separatorb 0\Separatorb -240 \RED{\omega_b}^2\big)\;\log\!\left[4 \RED{\omega_b}\,f(1\Separator 0\Separatorb -6 \RED{\omega_b})/\mu_{r}^{2}\right]\nonumber\\
&\qquad\qquad +F\big(1\Separator -12 \RED{\omega_b}\Separatorb -4 \RED{\omega_b}\Separator -51 \RED{\omega_b}^2\Separatorb 12 \RED{\omega_b}^2\Separatorb 4 \RED{\omega_b}^2\big)\;\log\!\left[6\beta_4 \RED{\omega_b}^{2}/\mu_{r}^{2}\right]\nonumber\\
&\qquad\qquad +F\big(\!-\!9\Separator -34 \RED{\omega_b}\Separatorb 72 \RED{\omega_b}\Separator 2 \RED{\omega_b}^2\Separatorb 172 \RED{\omega_b}^2\Separatorb -132 \RED{\omega_b}^2\big)\; \log\!\left[2 \RED{\omega_b}\, f(1\Separator \RED{\omega_b}\Separatorb -4 \RED{\omega_b})/\mu_{r}^{2}\right]\nonumber\\
&\qquad\qquad +F\big(\!-\!29\Separator -20 \RED{\omega_b}\Separatorb -148 \RED{\omega_b}\Separator 0\Separatorb -40 \RED{\omega_b}^2\Separatorb -68 \RED{\omega_b}^2\big)\; \log\!\left[-4 \RED{\omega_b}\, f(1\Separator 0\Separatorb 2 \RED{\omega_b})/\mu_{r}^{2}\right]\nonumber\\
&\qquad\qquad +F\big(6\Separator 66 \RED{\omega_b}\Separatorb 156 \RED{\omega_b}\Separator 39 \RED{\omega_b}^2\Separatorb -24 \RED{\omega_b}^2\Separatorb 96 \RED{\omega_b}^2\big)\;\log\!\left[-2 \RED{\omega_b}\, f(1\Separator - \RED{\omega_b}\Separatorb 0)/\mu_{r}^{2}\right]\nonumber\\
&\qquad\qquad +\text{IR divergent log terms}\Big),\label{eq:scalar-flipped-su5}
\end{align}
where the $F$ function is defined in eq.~\eqref{eq:define-functions-1-end}.
Let us note that the ``log-to-polynomial metamorphosis'' is less frequent here than in the previous case because only few (exclusively scalar) log prefactors blow up in the ``flipped'' SU(5) limit.
In fact, only the $O_{\IND{j}}$, $O_{\IND{l_2}}$, $O_{\IND{n_2}}$, $O_{\IND{o_2}}$, $T_{\IND{g}}$, $T_{\IND{o_2}}$ terms (see Appendix~\ref{app:full-one-loop-contributions}) contribute to the polynomial part of eq.~\eqref{eq:scalar-flipped-su5}.

\subsubsection{Exact Goldstone bosons\label{sect:exactG}}
There is another relatively cheap
consistency check of the method used for the calculation of the various $\Delta_{a,b}$ factors governing the leading one-loop contributions to the PGB masses of our interest: all would-be Goldstone modes associated to the gauge fields from the $SO(10)/SU(3)\times SU(2)\times U(1)$ coset should be massless to all orders in perturbation theory. Let us demonstrate that this is indeed the case at least for the most accessible
of these fields, namely, the $(3,2,-\tfrac{5}{6})+h.c.$ scalar which is  contained solely in the scalar $45$ of the $SO(10)$ and, hence, no mixing with the scalar $126$ needs to be considered.

We have indeed checked that its one-loop mass $m^{2}_c$ remains massless even at the 1-loop level, i.e., that each of its  $\Delta_{c}$-contributions -- defined along the lines of eqs.~\eqref{eq:mass1-all-triplet} and~\eqref{eq:mass1-all-octet} -- individually vanishes: the log and the polynomial terms of the gauge contribution, as well as the polynomial terms and log terms of the scalar contribution. In this case, the polynomial parts from scalars $\Delta_c^{S_{FIN}[\textrm{poly}]}$ and $\Delta_c^{S_{INF}[\textrm{poly}]}$ vanish  separately.\footnote{If one expresses the scalar mass matrix in a reordered basis, in which $\mathbf{M_S}^2$ acquires a block diagonal form, where each block consists of states with degenerate mass, it is easy to see explicitly that $\Delta_c^{S_{INF}[\textrm{poly}]} = 0$. The block structure alone then allows us to avoid the computation of complicated analytical forms of eigenvectors (modulo those of Goldstone modes, which are relatively simple) and automatically discards the polynomial contribution to the 1-loop mass of the would-be Goldstone pair $(3,2,-\tfrac{5}{6}) + (\bar{3},2,-\tfrac{5}{6})$ from the nested commutator term.}

Moreover, the structure of all the $\Delta_c$ contributions is such that it admits for a simple numerical check even outside the ``analytic simplification domain'' \eqref{eq:G3211-limit-original} which indeed yields zero for all randomly chosen values of  $\sigma$, $\gamma_2$ and  $a_2$.
\subsubsection{Diagrammatics}
Last, but not least, it is relatively straightforward to calculate some of the the leading polynomial parts of the one-loop corrections to the masses of the $(8,1,0)$ and $(1,3,0)$ PGBs by means of the standard perturbative expansion.
This applies, in particular, to the $\tau^{2}$-proportional (i.e., $SO(10)$ invariant) contribution, cf. eqs.~\eqref{eq:1looptauterm1} and~\eqref{eq:1looptauterm2}; the other leading polynomial terms, namely, those proportional to $\RED{\omega_{r}}^{2}$, $\RED{\omega_{b}}^{2}$ and/or $\RED{\omega_{r}\omega_{b}}$ turn out to be easily accessible only in the simplified version of the minimal model with 16 instead of 126 in the Higgs sector. In what follows, we shall first comment briefly on the salient points of the  corresponding calculation in the sample $45\oplus16$ Higgs model and then turn our attention to the $45\oplus126$ scenario of our current interest.
\subparagraph{The method:}
In the simplest scalar theory context with a pair of scalar fields $\Phi$ and  $\phi$ with only the latter developing a VEV it is straightforward to show that the leading order one-loop contribution to the mass (squared) of $\Phi$ can be formally written as~\cite{Lukasdiploma}
\be\label{eq:generic}
\Delta_{\Phi}=
\raisebox{-2mm}{\includegraphics[width=1.6cm]{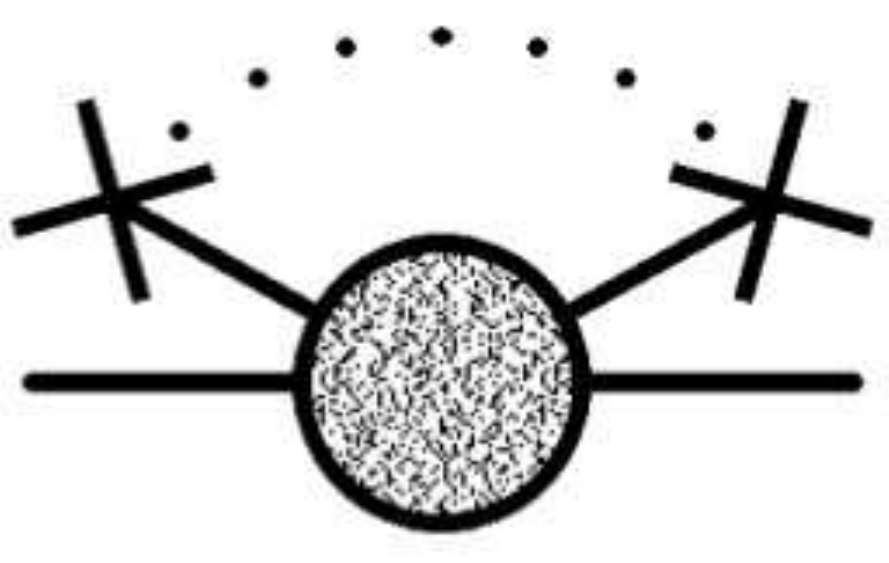}}\;\;-\frac{1}{\vev\phi}\;\;\raisebox{-2mm}{\includegraphics[width=1.6cm]{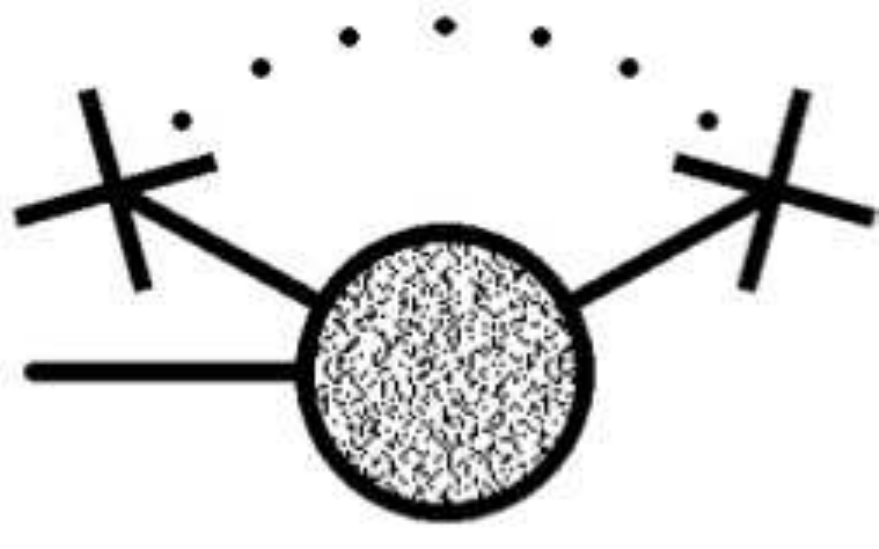}}\,,
\ee
where the graphs denote the sums of the one-loop contributions to the two-point and one-point functions with appropriate external legs $\Phi$, respectively,  while the dots between the crossed lines correspond to all possible insertions\footnote{Obviously, the calculation is performed in the unbroken phase formalism in which the VEV is kept in the interaction part of the Lagrangian.} of the VEVs of $\phi$. In the classical ``$\lambda\varphi^{4}$ context'' these structures can be formally expanded as
\bea
&\text{1-point:} &\raisebox{-2mm}{\includegraphics[width=6cm]{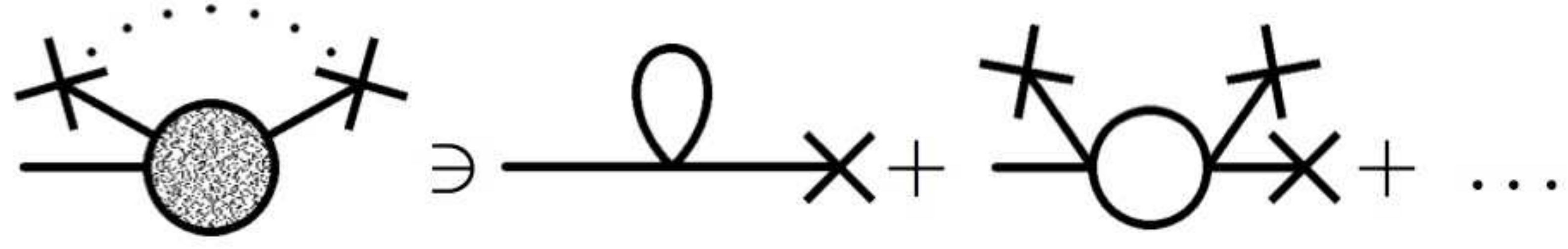}}\;,\label{eq:1point}
\\
&\text{2-point:}&\raisebox{-3mm}{\includegraphics[width=7cm]{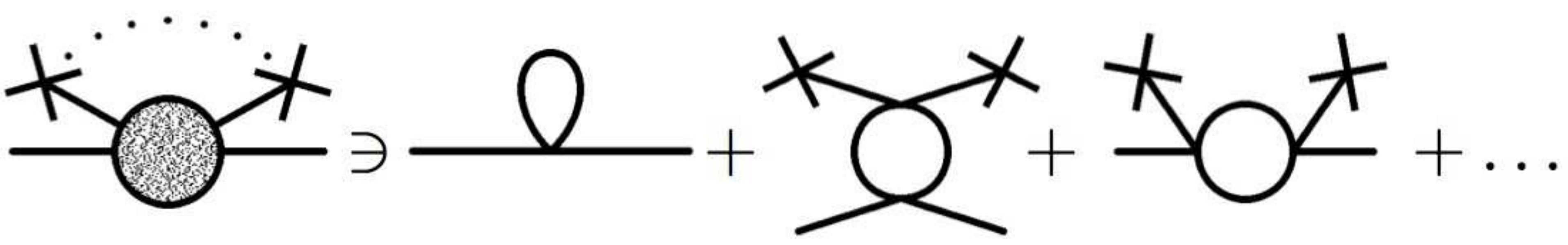}}\;,\label{eq:2point}
\eea
where the symbols with ``empty'' blobs stand for the usual Feynman diagrams of a given topology.
There are a few points worth making here:
\begin{enumerate}
\item Only some of the one-loop topologies above will generate a polynomial contribution to  $\Delta_{\Phi}$; for example, the first two displayed contributions to the 2-point function~\eqref{eq:2point} yield a polynomial contribution, while the third does not.
\item The undisplayed terms denoted by the ellipses above correspond to  the graphs with higher and higher number of insertions of the (pairs of) VEVs and, as such, they may generate a power-series-like structure of similar polynomial contributions; if the interactions are simple enough, the quotients of such power series may be identified and the series themselves may be eventually summed up in a closed form.
\item Note that if there is simultaneously a trilinear vertex at play, many more topologies become available; this will lead to ``mixed'' contributions proportional not only to the VEVs but also to the dimensionful trilinear vertex coupling (such as $\tau$ in the $SO(10)$ context of our interest). However, as long as one is interested in either the pure VEV-squared or the pure $\tau^{2}$ contribution to $\Delta_{\Phi}$, it is sufficient to focus on the relevant sub-series with only one kind of interactions (trilinear or quartic, respectively) connecting the VEV legs to the main loop.
\end{enumerate}

\subparagraph{Diagrammatics in the $45\oplus16$ scenario, the $\beta^{2}$-proportional polynomial piece:}
This all said, let us first turn our attention to the $SO(10)$ Higgs model featuring a simplified set of scalars transforming as  $45\oplus16$, see, e.g., reference~\cite{Bertolini:2009es}.  There are again three convenient limits in this setting corresponding to the assumed single VEV situation, namely
$\RED{\omega_r} = 0$ (the $3_{c}2_{L}2_{R}1_{B-L}$ limit), $\RED{\omega_{b}} = 0$ (the $4_{c}2_{L}1_{R}$ limit) and $\RED{\omega_r} = -\RED{\omega_{b}}$ (the flipped $SU(5)$ limit), in which the formalism above may be quite easily applied and clusters of graphs with contributions behaving as a power series identified in the 1-point and 2-point Green's function expansions.

For instance, focusing solely at the quartic interaction governed by the $\beta$ coupling (see~\cite{Bertolini:2009es} for its structure), the power-like behaviour of the polynomial contributions of the RHS of eqs.~\eqref{eq:1point} and~\eqref{eq:2point} may be readily inferred. The combinations~\eqref{eq:generic} of the sums of the corresponding power series are given in Table~\ref{Table:diagrammatics}.
\begin{table}[t]
\caption{\label{Table:diagrammatics}$\beta^{2}$-proportional parts of the polynomial corrections to the masses of the pseudo-Goldstone bosons in the simplified $45\oplus16$ scenario in various limits.}
\vskip 0.2cm
\centering
\begin{tabular}{cccc}
\toprule
$4\pi^{2}\Delta m^{2}_{\Phi}$ & $\quad$ $\RED{\omega_r} = 0$ $\quad$ & $\quad$ $\RED{\omega_{b}} = 0$ $\quad$ & $\quad$ $\RED{\omega_r} = -\RED{\omega_{b}}\equiv\RED{\omega}$ $\quad$\\
\addlinespace[\TABSPACE]\midrule
$\Phi=(8,1,0)$ & $3 \beta^{2}\RED{\omega_{b}}^{2}$& $ \beta^{2}\RED{\omega_{r}}^{2}$ & $5 \beta^{2}\RED{\omega}^{2}$\\
$\Phi=(1,3,0)$ & $2 \beta^{2}\RED{\omega_{b}}^{2}$& $ 2\beta^{2}\RED{\omega_{r}}^{2}$ & $5 \beta^{2}\RED{\omega}^{2}$\\
\hline
\end{tabular}
\end{table}
Remarkably, the information thus obtained in the three different limits above is just enough to reconstruct all three coefficients $c_{i}^{\Phi}$ of the expected form of the $\beta^{2}$-proportional polynomial one-loop contribution to $\Delta_{\Phi}$, namely  $\Delta_{\Phi}=\beta^{2}(c_{1}^{\Phi}\RED{\omega_{r}}^{2}+c_{2}^{\Phi}\RED{\omega_{r}\omega_{b}}+c_{3}^{\Phi}\RED{\omega_{b}}^{2})$.

Putting all this together, the polynomial pieces of the scalar-loop-generated corrections to the masses of the pseudo-Goldstone bosons $(1,3,0)$  and $(8,1,0) $ in the simplified $45\oplus16$ scenario read (in the same notation as before)
\bea
\Delta^{S[\rm poly],\beta^{2}}_{a} = \frac{1}{4\pi^{2}}\beta^2 \left( 2\RED{\omega_r}^2 - \RED{\omega_r\omega_b} + 2\RED{\omega_b}^2 \right)\,,\quad
\Delta^{S[\rm poly],\beta^{2}}_{b} & = & \frac{1}{4\pi^{2}}\beta^2 \left(\RED{\omega_r}^2 - \RED{\omega_r\omega_b} + 3\RED{\omega_b}^2 \right)\,,
\eea
which, indeed, coincides with the results of the existing effective potential analysis~\cite{Bertolini:2009es}.
%
\subparagraph{Diagrammatics in the $45\oplus16$ scenario, the $\tau^{2}$-proportional polynomial piece:}
Since the interaction of our interest here is trilinear, there are no $\tau$-proportional polynomial contributions popping up from the two-point part~\eqref{eq:1point} of formula~\eqref{eq:generic} and, hence, it is sufficient to consider only the $\tau$-proportional tadpoles~\eqref{eq:1point}. The summation of the relevant parts of the corresponding power series yields a universal (i.e., $SO(10)$ invariant) contribution
\bea
\Delta^{S[\rm poly],\tau^{2}}_{a}& = & \Delta^{S[\rm poly],\tau^{2}}_{b} = \frac{\tau^{2}}{4\pi^2}\,,
\eea
which, as before, coincides with that obtained in the effective potential approach~\cite{Bertolini:2009es}. Hence, at least for the one-loop polynomial corrections to the PGB triplet and octet masses, the purely diagrammatic approach admits an efficient cross-check of the EP results.
\subparagraph{Diagrammatics in the $45\oplus126$ scenario, the $\tau^{2}$-proportional polynomial piece:}
Finally, let us attempt to evaluate some of the $\Delta^{S[\rm poly]}_{a}$ and $\Delta^{S[\rm poly]}_{b}$ terms in the $45\oplus126$ model of our main interest.
Unfortunately, the presence of two types of quartic self-interactions between a pair of $45$'s and two $126$'s, i.e., the $\beta_{4}$ and $\beta_{4}'$ terms in the scalar potential~\eqref{eq:full-potential}, complicates the combinatorics of the VEV insertions in diagrams~\eqref{eq:generic} to such an extent that here we have managed to calculate just the universal $\tau^{2}$-proportional factor
\bea
\Delta^{S[\rm poly],\tau^{2}}_{a}& = & \Delta^{S[\rm poly],\tau^{2}}_{b} = \frac{35\tau^{2}}{8\pi^2}\,,
\eea
which, indeed, is identical to that obtained by the effective potential method earlier in this study, cf.~Sect.~\ref{sect:results} and formulae~\eqref{eq:1looptauterm1} and~\eqref{eq:1looptauterm2}.

Finally, it may be interesting to note that the parametrically higher level of complexity encountered in the $45\oplus126$ model in the relevant Feynman's diagram combinatorics is reflected  in the effective potential approach by the behaviour of the nested commutators~\eqref{eq:nested-commutator-series}: the interplay between $\beta_{4}$ and $\beta_{4}'$ was indeed  the main obstacle to writing the results of the EP calculation of Sect.~\ref{sect:results} in a more compact form as it could have been done if only one of these couplings was non-zero. In that respect, the simplicity of the diagrammatic  calculation in the $16\oplus45$ case can be attributed to the vanishing of the relevant nested commutators in the EP approach to this model.

\subsection{Viability of the minimal $SO(10)$ Higgs model at the one-loop level}
With all this information at hand one can finally re-address the central question -- whether the radiative corrections can really provide a regularization of the unpleasant tachyonic instabilities of the tree-level scalar spectrum.

Clearly, the first condition to be fulfilled is that there should exist a domain in the parameter space where the loop contributions to the masses of the pseudo-Goldstone triplet and octet scalars, cf.~\eqref{eq:mass1-all-triplet}--\eqref{eq:mass1-all-octet}, are large enough to compete with the problematic tree-level expressions therein and where all the other scalar-sector masses-squares are positive.

The former assumption is obviously attained in the regime when $a_{2}$ is sufficiently small. Remarkably enough, this is not only an option that one may choose at will, but rather a crucial ingredient of any perturbative account, see inequality~\eqref{eq:perturbativity}.  In this respect, {\em in the perturbative regime the quantum corrections always tend to regularize the notorious tachyonic instabilities of the scalar spectrum.}
As for the latter, there seems to be no simple analytic parametrization of the non-tachyonicity domain for the rest of the scalar spectrum. Hence, we provide a numerical example of a point in parameter space where the {\em entire scalar spectrum is regular and non-tachyonic}, see Table~\ref{tab:valid-spectrum}.
Note that the situation therein corresponds to the  simplified setting~\eqref{eq:G3211-limit-original} in  which $SO(10)$ gets first broken to $SU(3)_c \times \protect\linebreak[0]SU(2)_L \times \protect\linebreak[0]U(1)_R \times \protect\linebreak[0]U(1)_{B-L} \ $ whose subsequent breaking to the SM gauge group is arbitrarily delayed
(due to tiny $\sigma$) and, thus, there is an extra virtually massless mode corresponding to $m_{s_2}$ besides the usual massless would-be Goldstones \{$m_c$, $m_{p_1}$, $m_{q_1}$, $m_{r_1}$, $m_{s_1}$\}.

To this end, it is also worth noting that there is yet another field for which the radiative corrections may be important, namely, the lighter singlet scalar spanning predominantly on the upper-left $2\times 2$ block of matrix~\eqref{eq:mass-tree-end}. The point is that this state also becomes accidentally light in the potentially realistic ${a_{2}}\ll 1$ and $|\RED{\sigma}|\ll \text{max}\{|\RED{\omega_{r}}|,|\RED{\omega_{b}}|\}$ regime. However, unlike the two pseudo-Goldstones of our main interest here, this state still receives mass contributions from $\RED{\sigma}$ and, moreover, it is innocent from the gauge running perspective. Hence, we defer a thorough scrutiny of this issue to the future phenomenological analysis.
\begin{table}[t]
\caption{The shape of a potentially viable scalar spectrum (all masses in units of $\RED{\omega_{b}}$) at the leading order (corresponding to the one-loop level expressions~\eqref{eq:mass1-all-triplet}--\eqref{eq:mass1-all-octet} for the critical pseudo-Goldstones and tree-level formulae for the other scalars, respectively) computed in the limit~\eqref{eq:G3211-limit-original}. The underlying parameters were chosen as follows: ${a_2} = -{19.7}{|\sigma|^2}/{\omega_b^2}$ was determined by the vacuum stability condition~\eqref{eq:tau-tree-level} and the other parameters assumed the values $
{\RED{\omega_r}}/{\RED{\omega_b}} = 0.2$, ${\tau}/{\RED{\omega_b}} = -2$, ${\mu_r}/{\RED{\omega_b}} = 2$, $\beta_4 = \beta_4' = 0.4$, $a_0 = 0.2$ and $g = 0.7$. The scalar potential parameters $\alpha$, $\lambda_0$, $\lambda_2$, $\lambda_4$, $\lambda_4'$ and $\eta_2$ can remain unspecified since, at the leading order, they do not contribute to the above masses (as explained in Section~\ref{sect:results}). Note also that the loop-induced pseudo-Goldstone masses are much lighter than the bulk of the scalar spectrum which justifies the simple transition from the zero-momentum to the on-shell scheme advocated in Sect.~\ref{sect:mass-shell}. \label{tab:valid-spectrum}}
\vskip 0.2cm
\centering
\begin{tabular}{*{22}{c}|cc}
\toprule
${m_d}$ & ${m_e}$ & ${m_f}$ & ${m_g}$ & ${m_h}$ & ${m_i}$ & ${m_j}$ & ${m_k}$ & ${m_{l_1}}$ & ${m_{l_2}}$ & ${m_{m_1}}$ & ${m_{m_2}}$ & ${m_{n_1}}$ & ${m_{n_2}}$ & ${m_{o_1}}$ & ${m_{o_2}}$ & ${m_{o_3}}$ & ${m_{p_2}}$ & ${m_{q_2}}$ & ${m_{r_2}}$ & ${m_{r_3}}$ & ${m_{s_4}}$ & ${m_{a, \text{1-loop}}}$ & ${m_{b, \text{1-loop}}}$ \\
\addlinespace[\TABSPACE]\midrule
$1.9$ & $3.7$ & $3.3$ & $3.7$ & $3.0$ & $3.5$ & $3.6$ & $3.7$ & $3.6$ & $3.5$ & $2.6$ & $3.7$ & $3.3$ & $3.5$ & $3.7$ & $3.1$ & $3.2$ & $1.4$ & $2.9$ & $2.3$ & $3.7$ & $2.2$ & $0.27$ & $0.35$\\\addlinespace[\TABSPACE]
\bottomrule
\end{tabular}
\end{table}

\section{Conclusions and outlook\label{sect:Conclusions}}
In this work, we have calculated the one-loop corrections to the masses of a pair of scalar fields (transforming as $(1,3,0)$ and $(8,1,0$) under the SM gauge group) in the spectrum of the minimal non-supersymmetric $45\oplus 126$ $SO(10)$ Higgs model which, at the tree level, cause a notorious tachyonic instability in all of its potentially realistic vacuum configurations. The calculation confirms the former expectation made on qualitative grounds in~\cite{Bertolini:2009es} that the quantum effects can stabilize the phenomenologically viable vacua of the model at the one-loop level. Hence, the $45\oplus 126$ framework may be revived as a  basis of a full-fledged $SO(10)$ GUT construction that may be worth a further scrutiny concerning, namely, the fundamental signal of gauge unifications -- the proton lifetime. To this end, the current framework exhibits a particular robustness to various kinds of theoretical uncertainties, essentially unattainable in other popular GUT scenarios, which makes it very special when it comes to the exploitation of the information accessible in future megaton-scale experiments such as Hyper-Kamiokande or DUNE.

The consistency of the effective potential approach adopted in this study has been demonstrated by a number of explicit cross-checks of the results, including a thorough inspection of several of their enhanced-symmetry limits, a semi-analytic proof of the absence of a one-loop mass term for a selected would-be Goldstone mode, as well as a partial reconstruction of their purely polynomial parts by means of standard diagrammatic methods.

\subsection*{Future prospects}
Needless to say, a very detailed understanding of the behaviour of the critical components of the scalar spectra of the minimal $SO(10)$ GUTs featuring $45\oplus 126$ in its scalar sector is vital for any future phenomenological analysis going beyond the first rather simplified attempts~\cite{Bertolini:2012im,Kolesova:2014mfa}. The obvious goal here is to provide {\em really robust} estimates of the attainable proton lifetime with at least the leading theoretical uncertainties well under control, hopefully even within the expected ``sensitivity improvement window'' of the upcoming facilities. For that sake, a detailed analysis of the unification constraints including a two-loop renormalization group evolution of the gauge couplings is a particularly important element which, as an input, among other things, requires exactly the information supplied by this study.

On the practical side, this will consist of an extensive numerical simulation which would go even beyond the limit~\eqref{eq:G3211-limit-original} in which the analytic results have been displayed in this work (this is a purely technical issue related to the paramount complexity of the full results which, however, are also available). This will also facilitate the calculation of the radiative corrections to the mass of the third member of the potentially dangerous pseudo-Goldstone boson family which, in the physically interesting $\RED{\sigma}<{\rm max}\{\RED{\omega_r},\RED{\omega_b}\}$ regime, can be identified among the SM singlets~\eqref{eq:mass-tree-end}. In spite of its practical irrelevance for the gauge running it may  still represent an extra source of tachyonic instabilities a decisive scrutiny of the minimal model under consideration should not neglect. This, however, is beyond the scope of the current work and will be elaborated on elsewhere.
\section*{Acknowledgments}
The work of MM is supported by the Marie-Curie Career Integration Grant within the 7th European Community Framework Programme FP7-PEOPLE-2011-CIG, contract number PCIG10-GA-2011-303565. Two of the authors (MM, TM) acknowledge support by the Joint Program of Project Based Personnel Exchange of the Czech Ministry of Education (M\v{S}MT CZ) and Deutscher Akademischer Austauschdienst (DAAD), project nr.~7AMB15DE001, and by the  Foundation for support of science and research ``Neuron'';
they are indebted to Werner Porod, Christoph Gross and Thomas Garratt for the warm hospitality, friendly discussions and illuminating comments during their stays at the University in Wuerzburg.
VS acknowledges partial financial support from the Swiss National Science Foundation, European Research Council (ERC Starting Grant, agreement n. 278234, ``NewDark'' project) and the Slovenian Research Agency. MM and VS would like to thank CETUP* (Center for Theoretical Underground Physics and Related Areas), for the support during the 2015 Summer Program. VS would also like to thank for hospitality during his visit to IPNP in Prague, as well as to St\'{e}phane Lavignac and Marco Cirelli for hospitality during his stay at CEA Saclay.
Last, but not least, we would like to thank Borut Bajc and Helena Kole\v{s}ov\'a for reading through several fragments of the manuscript and for discussions.
\appendix
\section{The tree-level spectrum\label{app:tree-spectrum}}
\subsection{Gauge bosons\label{app:treegaugebosons}}
\noi
The scalar sector of the model is spanned on a $45$-dimensional $2$-index antisymmetric real representation $\phi_{ij}$ and a $126$-dimensional $5$-index antisymmetric self-dual
complex representation $\Sigma_{ijklm}$, defined as
\begin{align}
\Sigma_{ijklm} & = \frac{1}{\sqrt{2}} \left (\phi_{ijklm} - \frac{i}{5!} \, \epsilon_{ijklmabcde} \, \phi_{abcde} \right ),
\end{align}
where $\phi_{ijklm}$ and $\epsilon_{ijklmabcde}$ are the $252$-dimensional $5$-index antisymmetric real tensor and the completely antisymmetric Levi-Civita tensor with the positive signature $\epsilon_{12345678910} = +1$, respectively.
Under the infinitesimal $SO(10)$ transformations these objects change as
\begin{align}
\phi_{ij} & \to \phi_{ij} + i \left[\varphi, \phi \right]_{ij}, \\
\Sigma_{ijklm} & \to \Sigma_{ijklm} + i \left( \varphi_{ia} \, \Sigma_{ajklm} + \varphi_{jb} \, \Sigma_{ibklm} + \varphi_{kc} \, \Sigma_{ijclm} + \varphi_{ld} \, \Sigma_{ijkdm} + \varphi_{me} \, \Sigma_{ijkle} \right),
\end{align}
where
\begin{align}
\varphi_{ij} & \equiv \tfrac{1}{2} \, \varphi_{\alpha\beta} \, (\hat T^{\alpha\beta})_{ij}\, ,
\end{align}
and $\varphi_{\alpha\beta}$ are the infinitesimal antisymmetric real parameters, while $\hat T^{\alpha\beta}$ are the generators in the fundamental representation of $SO(10)$. We use the following definition of the relevant generators:
\begin{align}
(\hat T^{\alpha\beta})_{ij} & \equiv -\frac{i}{\sqrt{2}} \left(\delta_{\alpha i}\delta_{\beta j} - \delta_{\alpha j}\delta_{\beta i} \right),
\end{align}
satisfying the $SO(10)$ algebra
\begin{align}
\left [\hat T^{\alpha\beta}, \hat T^{\gamma\delta} \right ] & = \frac{i}{\sqrt{2}} \left (\delta^{\alpha\gamma} \hat T^{\beta\delta} + \delta^{\beta\delta} \hat T^{\alpha\gamma} - \delta^{\beta\gamma} \hat T^{\alpha\delta} - \delta^{\alpha\delta} \hat T^{\beta\gamma} \right ).
\end{align}
Consequently they are then canonically normalized to Dynkin index $1$
\begin{align}
\Tr\left (\hat T^{\alpha\beta} \hat T^{\gamma\delta}\right ) & = \delta_{\alpha\gamma}\delta_{\beta\delta} - \delta_{\alpha\delta} \delta_{\beta\gamma}\, ,
\end{align}
and, hence, the corresponding gauge coupling follows the usual $SU(5)$/ Standard Model normalization convention. The Latin and the Greek indices 
all  run from 1 to 10, while the summation over repeating indices is assumed everywhere.
\\
\noi
The gauge bosons in the adjoint representation
\begin{align}
(A_{\mu})_{ij} & \equiv \tfrac{1}{2} \, A_{\mu}^{\alpha\beta} \, (\hat T^{\alpha\beta})_{ij}\, ,
\end{align}
then transform as
\begin{align}
(A_{\mu})_{ij} & \to (A_{\mu})_{ij} + i \left[\varphi , A_{\mu} \right]_{ij} + \frac{1}{g} (\partial_{\mu} \varphi_{ij})\, .
\end{align}
\noi
Their mass term
$
\mathcal{L} \supset \tfrac{1}{2} \, {\mathbf{M}_\mathbf{G}^2}_{(\alpha\beta)(\gamma\delta)} A_{\mu}^{(\alpha\beta)} A^{\mu(\gamma\delta)}
$
then originates from the kinetic terms for the scalar fields
\begin{align}
\mathcal{L}_{kin} & \supset \frac{1}{4} 
(D_{\mu}\phi_{ij})^{*} D^{\mu}\phi_{ij} + \frac{1}{
5!} 
(D_{\mu}\Sigma_{ijklm})^{*} D^{\mu}\Sigma_{ijklm}\, , 
\end{align}
where the covariant derivatives are defined as
\begin{align}
D_{\mu}\phi_{ij} & \equiv \partial_{\mu}\phi_{ij}-i \, g \left[A_{\mu} , \phi \right]_{ij}, \\
D_{\mu}\Sigma_{ijklm} & \equiv \partial_{\mu}\Sigma_{ijklm}-i \, g \left\{(A_{\mu})_{ia} \Sigma_{ajklm} + (A_{\mu})_{jb} \Sigma_{ibklm} + (A_{\mu})_{kc} \Sigma_{ijclm} + (A_{\mu})_{ld} \Sigma_{ijkdm} + (A_{\mu})_{me} \Sigma_{ijkle} \right\},
\end{align}
and the subscripts $(i_1\ldots i_n)$ stand for the ${10}\choose{n}$ ordered n-tuples of indices. For instance, $(\alpha\beta)$ and $(\gamma\delta)$ represent the $45$ ordered pairs of indices that form the basis in which the $45\times 45$-dimensional gauge boson mass matrix $\mathbf{M}_\mathbf{G}^2$ is expressed as
\begin{align}
\mathbf{M}_\mathbf{G}^2 {}_{(\alpha\beta)(\gamma\delta)} = \frac{g^2}{2}
\bigg \{ & \delta^{\alpha\gamma} \langle \Sigma^{*}_{\beta(jklm)} \rangle \langle \Sigma_{\delta(jklm)} \rangle + \delta^{\beta\delta} \langle \Sigma^{*}_{\alpha(jklm)} \rangle \langle \Sigma_{\gamma(jklm)} \rangle - \delta^{\alpha\delta} \langle \Sigma^{*}_{\beta(jklm)} \rangle \langle \Sigma_{\gamma(jklm)} \rangle - \\
& - \delta^{\beta\gamma} \langle \Sigma^{*}_{\alpha(jklm)} \rangle \langle \Sigma_{\delta(jklm)} \rangle + \langle \Sigma^{*}_{\alpha\delta(klm)} \rangle \langle \Sigma_{\beta\gamma(klm)} \rangle + \langle \Sigma^{*}_{\beta\gamma(klm)} \rangle \langle \Sigma_{\alpha\delta(klm)} \rangle - \nonumber \\
& - \langle \Sigma^{*}_{\alpha\gamma(klm)} \rangle \langle \Sigma_{\beta\delta(klm)} \rangle - \langle \Sigma^{*}_{\beta\delta(klm)} \rangle \langle \Sigma_{\alpha\gamma(klm)} \rangle + \frac{1}{2} \left[\hat T^{(\alpha\beta)},\langle \phi \rangle \right]_{ij} \left[\hat T^{(\gamma\delta)}, \langle \phi \rangle \right]_{ji} \bigg \}+ \left(\begin{array}{c}\alpha\leftrightarrow \gamma\\\beta\leftrightarrow\delta\end{array}\right). \nonumber
\end{align}
Except for the $2\times 2$ block of singlets, the matrix $\mathbf{M}_\mathbf{G}^2|_{v}$ is already diagonal, which greatly simplifies the computation and final form of the nested commutator and logarithmic contributions to the scalar masses at 1 loop.
Its eigenvalues, representing the actual physical masses of the gauge bosons, are collected in Table~\ref{tab:gauge-boson-tree-masses}.
\begin{table}[t]
\caption{SM representations of gauge bosons, using definitions from Table~\ref{tab:all-SM-reps}. There are two kinds of contributions to the tree level masses of gauge bosons: $\Delta^{\phi}_{\mathbf{M}_\mathbf{G}^2}$ coming from the kinetic term of the 45-dimensional Higgs (due to the VEVs of the SM singlets $\sqrt{3} \, \omega_{b}$ and $\sqrt{2} \, \omega_{r}$) and $\Delta^{\Sigma}_{\mathbf{M}_\mathbf{G}^2}$ from the kinetic term of the 126-dimensional representations (due to $\sqrt{2} \, \sigma$ and $\sqrt{2} \, \sigma^*$). The final mass $\mathbf{M}_\mathbf{G}^2$ is the sum of both, producing $12$ massless gauge bosons, corresponding to $12$ unbroken generators of the SM; the dagger $\dagger$ denotes the presence of a broken generator and consequently a massive gauge boson. We kept both contributions $\Delta^{\phi}_{\mathbf{M}_\mathbf{G}^2}$ and $\Delta^{\Sigma}_{\mathbf{M}_\mathbf{G}^2}$ separate only to make various limits more evident. For that reason we also indicated from which representations of the Pati-Salam $G_{422}$ gauge group the individual SM representations descend. It is then easy to check that for $|\sigma| = 0$ in the $SU(5) \times U(1)_{Z}$ ($\omega_{r} = \omega_{b}$), flipped $SU(5)' \times U(1)_{Z'}$ ($\omega_{r} = -\omega_{b}$), left-right $G_{3221}$ ($\omega_r = 0$), $G_{421}$ ($\omega_{b} = 0$) and $G_{3211}$ ($\omega_{r}, \omega_{b} \neq 0$) limits , all the states within the same multiplets have the same mass and we get $25$, $25$, $15$, $19$ and $13$ massless gauge bosons, respectively. The two singlets $(1,1,0)$ in different Pati-Salam representations mix; the two given expressions are the two eigenvalues of the corresponding $2\times 2$ mass matrix.
\label{tab:gauge-boson-tree-masses}}
\vskip 0.2cm
\centering
\begin{tabular}{lrrr@{\hskip 1cm}lcc}
\toprule
\makebox[1.5cm][l]{$R\sim G_{321}$}&\makebox[1cm][r]{$\mathbb{R}/\mathbb{C}$}&\makebox[1cm][r]{$\#^{\phantom{\ast}}$}&\makebox[1cm][r]{size}&$R\;\subseteq G_{422}$&\makebox[3cm][c]{$\Delta^{\phi}_{\mathbf{M}_\mathbf{G}^2}$}& \makebox[2cm][c]{$\Delta^{\Sigma}_{\mathbf{M}_\mathbf{G}^2}$}\\
\midrule
$(8,1,0)$&$\mathbb{R}$&$1^{\phantom{\dagger}}$&$8$&$(15,1,1)$&$0$&$0$\\\addlinespace[\TABSPACE]
$(1,3,0)$&$\mathbb{R}$&$1^{\phantom{\dagger}}$&$3$&$(1,3,1)$&$0$&$0$\\\addlinespace[\TABSPACE]
$(1,1,0)$&$\mathbb{R}$&$2^\dagger$&$1$&$(15,1,1)$, $(1,1,3)$&$0;\;0$&$10\, g^2 |\RED{\sigma}|^2;\;0$\\\addlinespace[\TABSPACE]
$(1,1,+1)$&$\mathbb{C}$&$1^\dagger$&$2$&$(1,1,3)$&$2 g^2 \RED{\omega_{r}}^2$ & $\phantom{0}2\, g^2 |\RED{\sigma}|^2$\\\addlinespace[\TABSPACE]
$(3,1,+\tfrac{2}{3})$&$\mathbb{C}$&$1^\dagger$&$6$&$(15,1,1)$&$2 g^2 \RED{\omega_{b}}^2$ & $\phantom{0}2\, g^2 |\RED{\sigma}|^2$ \\\addlinespace[\TABSPACE]
$(3,2,+\tfrac{1}{6})$&$\mathbb{C}$&$1^\dagger$&$12$&$(6,2,2)$&$\tfrac{1}{2} \, g^2 (\RED{\omega_{r}}+\RED{\omega_{b}})^2$ & $\phantom{0}2\, g^2 |\RED{\sigma}|^2$\\\addlinespace[\TABSPACE]
$(3,2,-\tfrac{5}{6})$&$\mathbb{C}$&$1^\dagger$&$12$&$(6,2,2)$&$\tfrac{1}{2}\, g^2 (\RED{\omega_{r}}-\RED{\omega_{b}})^2$ & $0$\\\addlinespace[\TABSPACE]
\bottomrule
\end{tabular}
\end{table}
The $12$ zero modes correspond to the $12$ generators of $G_{321}$. The tree-level masses of the GUT-scale gauge bosons are
\begin{align}
M_{G}^2(3,2,-\tfrac{5}{6})&=\phantom{0}\tfrac{1}{2} \, g^2\, (\RED{\omega_{r}}-\RED{\omega_{b}})^2,\label{eq:gauge-boson-mass-begin}\\
M_{G}^2(3,2,+\tfrac{1}{6})&=\phantom{0}2 \, g^2 \,\left(|\RED{\sigma}|^2+\tfrac{1}{4} \, \left(\RED{\omega_{r}}+\RED{\omega_{b}}\right)^2\right),\\
M_{G}^2(3,1,+\tfrac{2}{3})&=\phantom{0}2 \, g^2\, (|\RED{\sigma}|^2+\RED{\omega_{b}}^2),\\
M_{G}^2(1,1,+1)&=\phantom{0}2 \, g^2\,(|\RED{\sigma}|^2+\RED{\omega_{r}}^2),\\
M_{G}^2(1,1,0)&=10 \, g^2\,|\RED{\sigma}|^2.\label{eq:gauge-boson-mass-end}
\end{align}

\subsection{Scalars\label{app:treespectrum}}

\noi
The tree level mass matrices of the various scalar fields belonging to $\phi$, $\Sigma$ and $\Sigma^*$ representations (whose decompositions and possible mixings are explained in Table~\ref{tab:all-SM-reps}; the states for a given SM rep.~$R$ in the table correspond to basis states for rows in the mass matrices, all in the same order, while columns have a basis of conjugate states $\overline{R}$) are
\begin{align}
M^{2}_{\IND{a}}&\equiv  M^2_S(1,3,0)=+2 \PARtm{a_2} \RED{\omega'} w^{1}_{[1,2]},\label{eq:mass-tree-begin}\\
M^{2}_{\IND{b}}&\equiv  M^2_S(8,1,0)=-2 \PARtm{a_2} \RED{\omega'} w^{1}_{[2,1]},\\
M^{2}_{\IND{c}}&\equiv  M^2_S(3,2,-\tfrac{5}{6})=0,\\
M^{2}_{\IND{d}}&\equiv  M^2_S(1,1,+2)= 4\, f(-\RED{\omega_r}\Separator 0\Separatorb 6 \RED{\omega_b} \RED{\omega_r})+8 s^{2}_{[1,1,4]},\\
M^{2}_{\IND{e}}&\equiv  M^2_S(1,3,-1)= 2\, f(- w^{1}_{[3,1]}\Separator 0\Separatorb 2 \RED{\omega_r} w^{1}_{[3,1]})+8 s^{2}_{[2,3,2]},\\
M^{2}_{\IND{f}}&\equiv  M^2_S(\bar{3},1,+\tfrac{4}{3})= 2\, f(-w^{1}_{[1,2]}\Separator \RED{\omega_b}^2\Separatorb 4 \RED{\omega_b} w^{1}_{[1,2]})+4 s^{2}_{[3,3,4]},\\
M^{2}_{\IND{g}}&\equiv  M^2_S(3,3,-\tfrac{1}{3})= 2\, f(-w^{1}_{[2,1]}\Separator \RED{\omega_b}^2\Separatorb 2 \RED{\omega}  w^{1}_{[2,1]})+4 s^{2}_{[3,3,4]},\\
M^{2}_{\IND{h}}&\equiv  M^2_S(6,3,+\tfrac{1}{3})= 2\, f(-\RED{\omega} \Separator 0\Separatorb 2 \RED{\omega}  w^{1}_{[2,1]})+8 s^{2}_{[1,1,4]},\\
M^{2}_{\IND{i}}&\equiv  M^2_S(\bar{6},1,-\tfrac{4}{3})= 4\, f(-\RED{\omega_b}\Separator 0\Separatorb 2 \RED{\omega_b} w^{1}_{[1,2]})+8 s^{2}_{[1,1,4]},\\
M^{2}_{\IND{j}}&\equiv  M^2_S(\bar{6},1,-\tfrac{1}{3})= 2\, f(-w^{1}_{[2,1]}\Separator \RED{\omega_r}^2\Separatorb 2 \RED{\omega}  w^{1}_{[2,1]})+4 s^{2}_{[3,3,4]},\\
M^{2}_{\IND{k}}&\equiv  M^2_S(\bar{6},1,+\tfrac{2}{3})= 4\, f(-\RED{\omega} \Separator 0\Separatorb 2 \RED{\omega}  \RED{\omega_b})+8 s^{2}_{[2,3,2]}.
\end{align}
\begin{align}
M^{2}_{\IND{l}} &\equiv  M^2_S(1,2,+\tfrac{1}{2})=
    \left(\begin{smallmatrix}
    f(-3 \RED{\omega} \Separator\tfrac{1}{2} w^{2}_{[7,-4,1]}\Separatorb 3 \RED{\omega}  w^{1}_{[3,1]})+4 s^{2}_{[3,3,4]} & -2 \PARtm{\gamma_{2}} \RED{\omega}  \RED{\omega'} \\
    -2 \PARtm{\gamma^{\ast}_{2}} \RED{\omega}  \RED{\omega'} & f(-w^{1}_{[3,1]}\Separator \tfrac{1}{2} w^{2}_{[7,4,1]}\Separatorb 3 \RED{\omega} w^{1}_{[3,1]})+8 s^{2}_{[1,1,-2]} \\
    \end{smallmatrix}\right),\\
M^{2}_{\IND{m}}&\equiv  M^2_S(3,2,+\tfrac{7}{6})=
    \left(\begin{smallmatrix}
    f(-w^{1}_{[1,3]}\Separator \tfrac{1}{2} \RED{\omega'}^2\Separatorb w^{1}_{[1,3]} w^{1}_{[5,1]})+8 s^{2}_{[1,1,4]} & -2 \PARtm{\gamma_{2}} \RED{\omega}  \RED{\omega'} \\
    -2 \PARtm{\gamma^{\ast}_{2}} \RED{\omega}  \RED{\omega'} & f(-w^{1}_{[5,1]}\Separator \tfrac{1}{2} \RED{\omega}^2\Separatorb w^{1}_{[1,3]} w^{1}_{[5,1]})+4 s^{2}_{[3,3,4]} \\
    \end{smallmatrix}\right),
\\
M^{2}_{\IND{n}}&\equiv  M^2_S(8,2,+\tfrac{1}{2})=
    \left(\begin{smallmatrix}
    f(-3 \RED{\omega} \Separator \tfrac{1}{2} \RED{\omega}^2\Separatorb 3 \RED{\omega}  w^{1}_{[3,1]})+4 s^{2}_{[3,3,4]} & -2 \PARtm{\gamma_{2}} \RED{\omega}  \RED{\omega'} \\
    -2 \PARtm{\gamma^{\ast}_{2}} \RED{\omega}  \RED{\omega'} & f(-w^{1}_{[3,1]}\Separator \tfrac{1}{2} \RED{\omega'}^2\Separatorb 3 \RED{\omega} w^{1}_{[3,1]})+8 s^{2}_{[1,1,4]} \\
    \end{smallmatrix}\right),
\end{align}
\begin{align}
M^{2}_{\IND{o}}&\equiv  M^2_S(\bar{3},1,+\tfrac{1}{3})=\nonumber\\
    &\left(\begin{smallmatrix}
    f(-2 \RED{\omega} \Separator w^{2}_{[1,0,1]}\Separatorb 4 \RED{\omega}  w^{1}_{[2,1]})+8 s^{2}_{[1,1,2]} & -4 \PARtm{\gamma^{\ast}_{2}} \RED{\omega} \RED{\omega'} & 2 \sqrt{2} (\PARtm{\beta_{4}} \RED{\omega_b} \RED{\omega_r}-8 s^{2}_{[0,0,1]}) \\
    -4 \PARtm{\gamma_{2}} \RED{\omega}  \RED{\omega'} & f(-2 w^{1}_{[2,1]}\Separator w^{2}_{[1,0,1]}\Separatorb 4 \RED{\omega} w^{1}_{[2,1]})+4 s^{2}_{[3,3,4]} & 0 \\
    2 \sqrt{2} (\PARtm{\beta_{4}} \RED{\omega_b} \RED{\omega_r}-8 s^{2}_{[0,0,1]}) & 0 & 2 f(-\RED{\omega}\Separator w^{2}_{[1,0,1]}\Separatorb 2 \RED{\omega}  w^{1}_{[2,1]})+8 s^{2}_{[1,1,0]} \\
    \end{smallmatrix}\right).\label{eq:Mo}\\
M^{2}_{\IND{p}}&\equiv  M^2_S(1,1,+1)=
    \left(\begin{smallmatrix}
     -2 \PARtm{a_{2}} \RED{\omega}  \RED{\omega_b}+2 s'^{2}_{[1,-2]} & 2\RED{\sigma}  f(1\Separator -\RED{\omega_r}\Separatorb -2 w^{1}_{[3,1]}) \\
    2 \RED{\sigma}^{\ast} f(1\Separator -\RED{\omega_r}\Separatorb -2 w^{1}_{[3,1]}) & -2 \RED{\omega_r} f(1\Separator -\RED{\omega_r}\Separatorb -2 w^{1}_{[3,1]}) \\
    \end{smallmatrix}\right),\\
M^{2}_{\IND{q}}&\equiv  M^2_S(\bar{3},1,-\tfrac{2}{3})=
    \left(\begin{smallmatrix}
    -2 \PARtm{a_{2}} \RED{\omega}  \RED{\omega_r}+2 s'^{2}_{[1,-2]} & -2 \RED{\sigma}^{\ast} f(1\Separator -\RED{\omega_b}\Separatorb -4 \RED{\omega} ) \\
    -2\RED{\sigma}  f(1\Separator -\RED{\omega_b}\Separatorb -4 \RED{\omega} ) & -2 \RED{\omega_b} f(1\Separator -\RED{\omega_b}\Separatorb -4 \RED{\omega}) \\
    \end{smallmatrix}\right),\\
M^{2}_{\IND{r}}&\equiv  M^2_S(3,2,+\tfrac{1}{6})=
    \left(\begin{smallmatrix}
    -4 \PARtm{a_{2}} \RED{\omega_b} \RED{\omega_r}+2 s'^{2}_{[1,-2]} &\RED{\sigma}  f(2\Separator -\RED{\omega}\Separatorb -2 w^{1}_{[5,3]}) & 4 \PARtm{\gamma_{2}}\RED{\sigma}  \RED{\omega'} \label{eq:Mrmatrix}\\
    \RED{\sigma}^{\ast} f(2\Separator -\RED{\omega}\Separatorb -2 w^{1}_{[5,3]}) & -\tfrac{1}{2} \RED{\omega}  f(2\Separator -\RED{\omega}\Separatorb -2 w^{1}_{[5,3]}) & -2 \PARtm{\gamma_{2}} \RED{\omega}  \RED{\omega'} \\
    4 \PARtm{\gamma^{\ast}_{2}} \RED{\sigma}^{\ast} \RED{\omega'} & -2 \PARtm{\gamma^{\ast}_{2}} \RED{\omega}  \RED{\omega'} & f(-w^{1}_{[5,3]}\Separator\tfrac{1}{2}\RED{\omega'}^2\Separatorb \RED{\omega}  w^{1}_{[5,3]})+8 s^{2}_{[2,3,2]} \\
    \end{smallmatrix}\right),\\
M^{2}_{\IND{s}}&\equiv  M^2_S(1,1,0)=\nonumber\\
    &\left(\begin{smallmatrix}
    24 \PARtm{a_{0}} \RED{\omega_b}^2-2 \PARtm{a_{2}} \RED{\omega'} w^{1}_{[2,1]}-12 s'^{2}_{[0,1]} & -\sqrt{6} (-8 \PARtm{a_{0}} \RED{\omega_b} \RED{\omega_r}+4 s'^{2}_{[0,1]}) & \sqrt{6}\RED{\sigma}  (2 \PARtm{\alpha}  \RED{\omega_b}+f(1\Separator 0\Separatorb -4 \RED{\omega} )) & \sqrt{6} \RED{\sigma}^{\ast} (2 \PARtm{\alpha}  \RED{\omega_b}+f(1\Separator 0\Separatorb -4 \RED{\omega} )) \\
    -\sqrt{6} (-8 \PARtm{a_0} \RED{\omega_b} \RED{\omega_r}+4 s'^{2}_{[0,1]}) & 16 \PARtm{a_{0}} \RED{\omega_r}^2+2 \PARtm{a_{2}} \RED{\omega'} w^{1}_{[1,2]}-8 s'^{2}_{[0,1]} & 2\RED{\sigma}  (2 \PARtm{\alpha}  \RED{\omega_r}+f(1\Separator 0\Separatorb -2 w^{1}_{[3,1]})) & 2 \RED{\sigma}^{\ast} (2 \PARtm{\alpha}  \RED{\omega_r}+f(1\Separator 0\Separatorb -2 w^{1}_{[3,1]})) \\
    \sqrt{6} \RED{\sigma}^{\ast} (2 \PARtm{\alpha}  \RED{\omega_b}+f(1\Separator 0\Separatorb -4 \RED{\omega} )) & 2 \RED{\sigma}^{\ast} (2 \PARtm{\alpha}  \RED{\omega_r}+f(1\Separator 0\Separatorb -2 w^{1}_{[3,1]})) & 4 \PARtm{\lambda_{0}}|\RED{\sigma}|^{2} & 4 \PARtm{\lambda_{0}} \RED{\sigma}^{\ast\,2} \\
    \sqrt{6}\RED{\sigma}^{\phantom{\ast}}  (2 \PARtm{\alpha}  \RED{\omega_b}+f(1\Separator 0\Separatorb -4 \RED{\omega} )) & 2\RED{\sigma}^{\phantom{\ast}}  (2 \PARtm{\alpha}  \RED{\omega_r}+f(1\Separator 0\Separatorb -2 w^{1}_{[3,1]})) & 4 \PARtm{\lambda_{0}}\RED{\sigma}^{2\,\phantom{\ast}} & 4 \PARtm{\lambda_{0}}|\RED{\sigma}|^{2} \\
    \end{smallmatrix}\right)\label{eq:mass-tree-end}.
\end{align}
\noi
In the expressions above (and in section~\ref{sec:checks-limits}) we have used the definitions
\begin{align}
w^{n}_{[a_n,\ldots,a_0]}&:=\sum_{i=0}^{n}a_{n-i}\,\RED{\omega_b}^{n-i}\,\RED{\omega_r}^{i},\label{eq:define-functions-1-begin}\\
\RED{\omega}&:=\phantom{-}\RED{\omega_b}+\RED{\omega_r}=w^{1}_{[1,1]},\\
\RED{\omega'}&:=-\RED{\omega_b}+\RED{\omega_r}=w^{1}_{[-1,1]},\\
s^{2}_{[a_1, a_2, a_3]}&:=(\PAR{\lambda_2}\,a_1+\PAR{\lambda_4}\,a_2+\PAR{\lambda_4'}\,a_3)\,|\RED{\sigma}|^2,\\
s'^{2}_{[a_1, a_2]}&:=(\PAR{\beta_4}\,a_1+\PAR{\beta_4'}\,a_2)\,|\RED{\sigma}|^{2},\\
f(z_1\Separator z_2\Separatorb z_3)&:=\PAR{\tau}\,z_1+\PAR{\beta_4}\,z_2+\PAR{\beta_4'}\,z_3,\\
F(z_1\Separator z_2\Separatorb z_3\Separator z_4\Separatorb z_5\Separatorb z_6)&:=\PAR{\tau}^2\,z_1+\PAR{\tau}(\PAR{\beta_4}\,z_2+\PAR{\beta_4'}\,z_3)+\PAR{\beta_4}^{2}\,z_4+\PAR{\beta_4}\PAR{\beta_4'}\,z_5+\PAR{\beta_4'}^{2}\,z_6.
\label{eq:define-functions-1-end}
\end{align}
\noi
The rationale behind this notation is the following:
\begin{enumerate}
\item Labels $a_i$ enclosed in square brackets in subscripts of functions $w^{n}$, $s^{2}$ and $s'^{2}$ are integers parametrising the coefficients of the corresponding polynomial expressions, $z_i$'s enclosed in round brackets are arguments of functions $f$ and $F$; note that not all of them have  the same mass dimension $d$ because $\tau$ is a $d=1$ parameter ($\beta_4$ and $\beta_4'$ are dimensionless). Variables with the same mass dimensions are separated with commas, while groups with different mass dimensions are separated with semicolons.
\item  Symbols $\RED{\omega}$ and $\RED{\omega'}$ denote the sum and the difference of the VEVs $\RED{\omega_b}$ and $\RED{\omega_r}$, respectively; these were introduced mainly for their relevance in the standard and/or the flipped $\mathrm{SU}(5)\times\mathrm{U}(1)$ limits.
\item In various limits of increased symmetry the RH sides of the equalities above simplify to
    \begin{align}
    \RED{\sigma}\to 0&\quad\Rightarrow\quad s^{2}_{[a_1,a_2,a_3]}\to 0,\; s'^{2}_{[a_1,a_2]}\to 0,\\
    \RED{\omega_{r}}\to 0&\quad\Rightarrow\quad w^{n}_{[a_n,\ldots,a_0]}\to a_n\,\RED{\omega_b}^{n}\\
    \RED{\omega_{b}}\to 0&\quad\Rightarrow\quad w^{n}_{[a_n,\ldots,a_0]}\to a_0\,\RED{\omega_r}^{n},\\
    \RED{\omega_{r}}\to +\RED{\omega_{b}}&\quad\Rightarrow\quad w^{n}_{[a_n,\ldots,a_0]}\to \sum_{i}a_{n-i}\,\RED{\omega_b}^{n},\\
    \RED{\omega_{r}}\to -\RED{\omega_{b}}&\quad\Rightarrow\quad w^{n}_{[a_n,\ldots,a_0]}\to \sum_i (-1)^{i}a_{n-i}\,\RED{\omega_b}^{n}.
    \end{align}
\end{enumerate}
For more details on these limits the reader is kindly referred to Table~\ref{tab:groups-in-limits} and the discussion in Section~\ref{sec:checks-limits}.
\vskip 2mm
\noi
A few other comments regarding the tree-level spectrum are worth making here:
\begin{enumerate}
\item
There are $3$ equations for the vacuum configuration of the 3 independent VEVs in the $45 \oplus 126$ model (let us reiterate that $\RED{\omega_{r}}$ and $\RED{\omega_{b}}$ are real while $\RED{\sigma}$ can be complex).
These quantities can be traded for the $3$ mass parameters ($\mu^2$, $\nu^2$, $\tau$) via the eqs.~\eqref{eq:mu-tree-level}--\eqref{eq:tau-tree-level}. For the sake of simplicity, we shall sometimes use a hybrid notation in which the $\tau$ parameter may appear in the relevant expressions alongside the three VEVs above, see eqs.~\eqref{eq:mass-tree-begin}--\eqref{eq:mass-tree-end}; however, in practice, the third condition of eq.~\eqref{eq:tau-tree-level} should always be used to eliminate~$\tau$.
\item
The tree level scalar spectrum of this model has been previously calculated in \cite{Bertolini:2012im}; our results agree with those given there with the  only exception of the numerical coefficients in front of $\lambda_4$ terms which have been corrected.
\item
With no special correlation among the three VEVs above, the $SO(10) \to G_{321}$ breaking is achieved and one should end up with $33$ massless would-be Goldstone modes identified with the broken $SO(10)$ generators. There is indeed a zero eigenvalue in each of the following mass matrices above: $M^{2}_{\IND{c}}$, $M^{2}_{\IND{p}}$, $M^{2}_{\IND{q}}$, $M^{2}_{\IND{r}}$ and $M^{2}_{\IND{s}}$, which together with the multiplicity of each of the states and their conjugated counterparts (as indicated in the column ``size'' of Table~\ref{tab:all-SM-reps}) gives the desired number of would-be Goldstone modes.
\end{enumerate}

\section{The one-loop contributions to the pseudo-Goldstone boson masses\label{app:full-one-loop-contributions}}
In this Appendix, we present the full mathematical form of the $\Delta$-terms entering the relevant formulae for the one-loop pseudo-Goldstone triplet $(1,3,0)$ and octet $(8,1,0)$ masses~\eqref{eq:mass1-all-triplet} and~\eqref{eq:mass1-all-octet}.
For that sake, it is convenient to define the symbols
\begin{align}
C=-\pi^2\,\RED{\omega'},\qquad
R^{2}=\sqrt{\RED{\omega_b}^4+34\,\RED{\omega_b}^2\RED{\omega_r}^{2}+\RED{\omega_r}^4}.\label{eq:Rsquared}
\end{align}
\subsection{Gauge boson contributions\label{app:gaugeloops}}
Here we present the 1-loop contributions to the scalar masses~\eqref{eq:mass1-all-triplet}--\eqref{eq:mass1-all-octet} coming from the gauge bosons running in the loops. All the terms in this subsection have been computed in full generality, i.e., they are valid for arbitrary values of all parameters.
\begin{align}
\Delta^{G[\text{poly}]}_{a}  = \frac{g^4}{{16} \pi ^2} w^2_{[19,1,16]}, \qquad
\Delta^{G[\text{poly}]}_{b} & = \frac{g^4}{{16} \pi ^2} w^2_{[22,1,13]},
\end{align}
\begin{align}
\Delta^{G[\text{log}]}_{a} &= \frac{3 g^4}{16 \, C} \Bigg( \Bigg.  -8 \RED{\omega_b} \left(|\RED{\sigma}|^2+\RED{\omega_b}^2\right) \log \left[\frac{{2} g^2 \left(|\RED{\sigma}|^2+\RED{\omega_b}^2\right)}{\mu_r^2}\right]+4 \RED{\omega_r} \left(|\RED{\sigma}|^2 + \RED{\omega_r}^2 \right) \log \left[\frac{{2} g^2 \left(|\RED{\sigma}|^2+\RED{\omega_r}^2\right)}{\mu_r^2}\right] + \nonumber \\
& \qquad \Bigg. +\left(4 |\RED{\sigma}|^2 w^1_{[2,-1]}+\RED{\omega}^{2} w^1_{[5,-4]}\right) \log \left[\frac{{2} g^2 \left(|\RED{\sigma}|^2+\tfrac{1}{4}\RED{\omega}^2\right)}{\mu_r^2}\right]
- 2 \RED{\omega'}^3 \log \left[\frac{\tfrac{1}{{2}} g^2 \RED{\omega'}^2}{\mu_r^2}\right] \Bigg ),\\
\Delta^{G[\text{log}]}_{b} &= \frac{3 g^4}{32 \, C} \Bigg( \Bigg.  -4 \left(|\RED{\sigma}|^2 w^1_{[3,1]}+\RED{\omega_b}^2 w^1_{[1,3]}\right) \log \left[\frac{{2} g^2 \left(|\RED{\sigma}|^2+\RED{\omega_b}^2\right)}{\mu_r^2}\right] +8 \RED{\omega_r} \left(|\RED{\sigma}|^2+\RED{\omega_r}^2\right) \log \left[\frac{{2} g^2 \left(|\RED{\sigma}|^2+\RED{\omega_r}^2\right)}{\mu_r^2}\right] + \nonumber \\
& \qquad \Bigg. +\left(4 |\RED{\sigma}|^2 w^1_{[3,-1]}+\RED{\omega}^{2} w^1_{[7,-5]}\right) \log \left[\frac{{2} g^2 \left(|\RED{\sigma}|^2+\tfrac{1}{4}\RED{\omega}^2\right)}{\mu_r^2}\right]
-\RED{\omega'}^3 \log \left[\frac{\tfrac{1}{{2}} g^2 \RED{\omega'}^2}{\mu_r^2}\right]\Bigg) .
%
%
\end{align}
\subsection{Scalar contributions\label{app:scalarloops}}
Here we list the scalar-loop-induced contributions to the masses~\eqref{eq:mass1-all-triplet}--\eqref{eq:mass1-all-octet}. The $\Delta^{S_{\text{FIN}}[\text{poly}]}$ structures are computed for arbitrary values of parameters, while $\Delta^{S_{\text{INF}}[\text{poly}]}$ and $\Delta^{S[\text{log}]}$ are only given in the limit~\eqref{eq:G3211-limit-original}.
The tree-level value of $\tau$ from eq.~\eqref{eq:tau-tree-level} should eventually be inserted into the expressions below.
\begin{align}
\Delta^{S_{\text{FIN}}[\text{poly}]}_{\IND{a}} & = -\frac{\RED{\omega'} w^1_{[1,2]}}{16 \pi ^2} \left(96
      a_0 a_2+76 a_2^2-5 (\beta_4-10 \beta_4') (\beta_4-2
      \beta_4')+560 |\gamma_2|^2\right),\\
\Delta^{S_{\text{FIN}}[\text{poly}]}_{\IND{b}} & = +\frac{\RED{\omega'} w^1_{[2,1]}}{16 \pi ^2} \left(96
   a_0 a_2+76 a_2^2-5 (\beta_4-10 \beta_4') (\beta_4-2
   \beta_4')+560 |\gamma_2|^2\right).
\end{align}
The polynomial parts of the nested commutator contributions computed in the limit~\eqref{eq:G3211-limit-original} are
\begin{align}
\Delta^{S_{\text{INF}}[\text{poly}]}_{\IND{a}} & = \frac{1}{\phantom{2}8\pi^{2}}\,F\Big(35\Separator 0\Separatorb 0\Separator 3 w^{2}_{[6,0,5]}\Separatorb 60 \RED{\omega_r}^{2}\Separatorb 60 w^{2}_{[4,0,1]}\Big),\label{eq:1looptauterm1}\\
\Delta^{S_{\text{INF}}[\text{poly}]}_{\IND{b}} & = \frac{1}{24\pi^{2}}\,F\Big(105\Separator 0\Separatorb 0\Separator 216 \RED{\omega_b}^2 \tfrac{\RED{\omega_r}^4}{R^4}+w^{2}_{[71,0,32]}\Separatorb 180 \RED{\omega_b}^2\Separatorb 60 w^{2}_{[7,0,8]}\Big),\label{eq:1looptauterm2}
\end{align}
and the log terms read
\begin{align}
\Delta^{S[\text{log}]}_{\IND{a}}  \equiv \sum_{x\in X} T_x \log(m_x^2/\mu_r^2), \qquad
\Delta^{S[\text{log}]}_{\IND{b}} \equiv \sum_{x\in X} O_x \log(m_x^2/\mu_r^2),
\label{eq:logs}
\end{align}
\noindent
where $x$ runs over the set of indices $
X=\{\IND{d},\IND{e},\IND{f},\IND{g},\IND{h},\IND{i},\IND{j},\IND{k},\IND{l_1},\IND{l_2},\IND{m_1},\IND{m_2},\IND{n_1},\IND{n_2},\IND{o_1},\IND{o_2},\IND{o_3},\IND{p},\IND{q},\IND{r_1},\IND{r_2}\}.
$
\vskip 2mm
\noi
The $T_x$ and $O_x$ prefactors are
\begin{align}
T_{\IND{d}}&= \tfrac{1}{2\,\RED{C}}\;F\Big( \RED{\omega_r} \Separator   0 \Separatorb   -4 \RED{\omega_r} w^{1}_{[3,-1]} \Separator   0 \Separatorb   0 \Separatorb   12 \RED{\omega_b} \RED{\omega_r} w^{1}_{[3,-2]} \Big), \\
T_{\IND{e}}&= \tfrac{1}{8\,\RED{C}}\;F\Big( -w^{1}_{[7,5]} \Separator   2 \RED{\omega'} w^{1}_{[3,1]} \Separatorb   -2 w^{2}_{[3,20,1]} \Separator   0 \Separatorb   -4 \RED{\omega'} \RED{\omega_r} w^{1}_{[3,1]} \Separatorb   4 w^{3}_{[18,-3,17,4]} \Big), \\
T_{\IND{f}}&= \tfrac{1}{2\,\RED{C}}\;F\Big( w^{1}_{[1,2]} \Separator   2 \RED{\omega_b} \RED{\omega_r} \Separatorb   -2 w^{1}_{[1,2]} w^{1}_{[3,-2]} \Separator   -\RED{\omega_b}^3 \Separatorb   -2 \RED{\omega_b}^2 w^{1}_{[1,6]} \Separatorb   8 \RED{\omega_b} w^{1}_{[1,-2]} w^{1}_{[1,2]} \Big), \\
T_{\IND{g}}&= \tfrac{1}{8\,\RED{C}}\;F\Big( -9 \RED{\omega_r} \Separator   \tfrac{3}{\RED{\omega}} w^{3}_{[-11,3,8,2]} \Separatorb   -18 w^{2}_{[-4,2,1]} \Separator -\tfrac{6}{\RED{\omega}} \RED{\omega_b}^2 \RED{\omega'} w^{1}_{[3,1]} \Separatorb   6 w^{3}_{[9,2,-8,-2]} \Separatorb   -12 w^{3}_{[4,3,-9,-4]} \Big), \\
T_{\IND{h}}&= \tfrac{1}{4\,\RED{C}}\;F\Big( 3 w^{1}_{[3,-1]} \Separator   6\RED{\omega}  \RED{\omega'} \Separatorb   -6 w^{2}_{[3,0,5]} \Separator   0 \Separatorb   -12\RED{\omega}  \RED{\omega'} w^{1}_{[2,1]} \Separatorb   -12 w^{3}_{[4,3,-9,-4]} \Big), \\
T_{\IND{i}}&= \tfrac{1}{\RED{C}}\;F\Big( -\RED{\omega_b} \Separator   0 \Separatorb   8 \RED{\omega_b} \RED{\omega_r} \Separator   0 \Separatorb   0 \Separatorb   4 \RED{\omega_b} w^{1}_{[1,-2]} w^{1}_{[1,2]} \Big), \\
T_{\IND{j}}&= \tfrac{1}{4\,\RED{C}}\;F\Big( -w^{1}_{[2,1]} \Separator   -\RED{\omega_r} w^{1}_{[12,5]} \Separatorb   2 \RED{\omega_r} w^{1}_{[2,1]} \Separator   6 \RED{\omega_r}^3 \Separatorb   2 \RED{\omega_r} w^{2}_{[12,19,6]} \Separatorb   4\RED{\omega}  \RED{\omega_b} w^{1}_{[2,1]} \Big), \\
T_{\IND{k}}&= \tfrac{1}{\RED{C}}\;F\Big(\RED{\omega}  \Separator   0 \Separatorb   4\RED{\omega}  \RED{\omega_r} \Separator   0 \Separatorb   0 \Separatorb   -4\RED{\omega}  \RED{\omega_b} w^{1}_{[1,2]} \Big), \\
T_{\IND{l_1}}&= \tfrac{1}{32\,\RED{C}}\;F\Big( 4 w^{1}_{[2,1]} \Separator   w^{2}_{[19,42,-13]} \Separatorb   -2 w^{2}_{[3,2,7]} \Separator   -\tfrac{1}{3} w^{3}_{[95,-48,3,2]} \Separatorb   -4 \RED{\omega_b} w^{2}_{[29,30,-5]} \Separatorb   -4 w^{3}_{[9,12,1,2]} \Big), \\
T_{\IND{l_2}}&= \tfrac{1}{16\,\RED{C}}\;F\Big( -2\RED{\omega}  \Separator   \tfrac{1}{6} w^{2}_{[27,58,23]} \Separatorb   w^{2}_{[15,10,7]} \Separator   -\tfrac{1}{6} w^{3}_{[11,48,39,10]} \Separatorb   -2 w^{3}_{[11,14,13,4]} \Separatorb   -2 w^{3}_{[9,12,1,2]} \Big), \\
T_{\IND{m_1}}&= \tfrac{1}{32\,\RED{C}}\;F\Big( 4 w^{1}_{[5,9]} \Separator   \RED{\omega'} w^{1}_{[15,-31]} \Separatorb   -2 w^{2}_{[81,134,9]} \Separator   -4 \RED{\omega'}^3 \Separatorb   -2 \RED{\omega'} w^{2}_{[25,-83,10]} \Separatorb   4 w^{3}_{[79,108,-1,-18]} \Big), \\
T_{\IND{m_2}}&= \tfrac{1}{32\,\RED{C}}\;F\Big( -4 w^{1}_{[16,5]} \Separator   w^{2}_{[-45,26,23]} \Separatorb   2 w^{2}_{[-51,94,41]} \Separator   2\RED{\omega} ^2 w^{1}_{[5,-4]} \Separatorb   2 w^{3}_{[7,84,7,-22]} \Separatorb   4 w^{3}_{[79,108,-1,-18]} \Big), \\
T_{\IND{n_1}}&= \tfrac{1}{2\,\RED{C}}\;F\Big( 2 w^{1}_{[2,1]} \Separator   -\tfrac{1}{2}\RED{\omega}  w^{1}_{[5,1]} \Separatorb   w^{2}_{[-3,-2,-7]} \Separator   \tfrac{1}{3}\RED{\omega} ^2 w^{1}_{[5,-4]} \Separatorb  \RED{\omega}  w^{2}_{[11,1,-6]} \Separatorb   -2 w^{3}_{[9,12,1,2]} \Big), \\
T_{\IND{n_2}}&= \tfrac{1}{12\,\RED{C}}\;F\Big( -12\RED{\omega}  \Separator   -\RED{\omega'} w^{1}_{[27,-11]} \Separatorb   6 w^{2}_{[15,10,7]} \Separator   -4 \RED{\omega'}^3 \Separatorb   6 \RED{\omega'} w^{2}_{[7,11,-2]} \Separatorb   -12 w^{3}_{[9,12,1,2]} \Big), \\
T_{\IND{o_1}}&= \tfrac{1}{32\,\RED{C}}\;F\Big( 4\RED{\omega}  \Separator   3 w^{2}_{[1,-2,-5]}-\tfrac{1}{R^2} w^{4}_{[-9,98,68,-62,13]} \Separatorb   -8\RED{\omega}  w^{1}_{[3,1]} \Separator  w^{3}_{[-7,36,-23,12]}-\tfrac{3}{R^2} w^{5}_{[3,-28,38,-64,19,-4]} \Separatorb\nonumber\\
    &\qquad\qquad   -6 w^{1}_{[3,4]} w^{2}_{[1,-2,-1]}-\tfrac{2}{R^2} w^{5}_{[19,-186,-200,90,37,-12]} \Separatorb   16\RED{\omega}  \RED{\omega_b} w^{1}_{[2,1]} \Big), \\
T_{\IND{o_2}}&= \tfrac{1}{16\,\RED{C}}\;F\Big( -\!2 w^{1}_{[2,1]} \Separator   \tfrac{1}{\RED{\omega}} w^{3}_{[93,43,-1,1]} \Separatorb   4 \RED{\omega_r} w^{1}_{[2,1]} \Separator   \tfrac{\RED{\omega_b}}{\RED{\omega}} w^{1}_{[-23,1]} w^{2}_{[1,0,1]} \Separatorb   -2 \RED{\omega_b} w^{2}_{[91,42,-3]} \Separatorb   8\RED{\omega}  \RED{\omega_b} w^{1}_{[2,1]} \Big), \\
T_{\IND{o_3}}&= \tfrac{1}{32\,\RED{C}}\;F\Big( 4\RED{\omega}  \Separator   3 w^{2}_{[1,-2,-5]}+\tfrac{1}{R^2} w^{4}_{[-9,98,68,-62,13]} \Separatorb   -8\RED{\omega}  w^{1}_{[3,1]} \Separator w^{3}_{[-7,36,-23,12]}+\tfrac{3}{R^2} w^{5}_{[3,-28,38,-64,19,-4]} \Separatorb\nonumber\\
    &\qquad\qquad  -6 w^{1}_{[3,4]} w^{2}_{[1,-2,-1]}+\tfrac{2}{R^2} w^{5}_{[19,-186,-200,90,37,-12]} \Separatorb   16\RED{\omega}  \RED{\omega_b}w^{1}_{[2,1]} \Big), \\
T_{\IND{p}}&= \tfrac{1}{8\,\RED{C}}\;F\Big( \RED{\omega_r} \Separator   -3 \RED{\omega_r}^2 \Separatorb   -2 \RED{\omega_r} w^{1}_{[6,1]} \Separator   2 \RED{\omega_r}^3 \Separatorb   2 \RED{\omega_r}^2 w^{1}_{[9,2]} \Separatorb   12 \RED{\omega_b} \RED{\omega_r} w^{1}_{[3,1]} \Big), \\
T_{\IND{q}}&= \tfrac{1}{4\,\RED{C}}\;F\Big( -\RED{\omega_b} \Separator   3 \RED{\omega_b}^2 \Separatorb   2 \RED{\omega_b} w^{1}_{[3,4]} \Separator   -2 \RED{\omega_b}^3 \Separatorb   -2 \RED{\omega_b}^2 w^{1}_{[5,6]} \Separatorb   -8\RED{\omega}  \RED{\omega_b} w^{1}_{[1,2]} \Big), \\
T_{\IND{r_1}}&= \tfrac{1}{32\,\RED{C}}\;F\Big( 4 w^{1}_{[2,-1]} \Separator   -3\RED{\omega}  w^{1}_{[7,-5]} \Separatorb   -2 w^{2}_{[27,2,-1]} \Separator   2\RED{\omega} ^2 w^{1}_{[5,-4]} \Separatorb   2\RED{\omega}  w^{2}_{[43,-7,-14]} \Separatorb   4 w^{3}_{[19,12,11,6]} \Big), \\
T_{\IND{r_2}}&= \tfrac{1}{32\,\RED{C}}\;F\Big( -4 w^{1}_{[1,3]} \Separator   -\RED{\omega'} w^{1}_{[9,23]} \Separatorb   2 w^{2}_{[39,-38,-33]} \Separator   -4 \RED{\omega'}^3 \Separatorb   2 \RED{\omega'} w^{2}_{[35,-1,-2]} \Separatorb   4 w^{3}_{[19,12,11,6]} \Big),\\
T_{\IND{s}}&=0\,,
\end{align}
\begin{align}
O_{\IND{d}}&= \tfrac{1}{2\,\RED{C}}\; F\Big( \RED{\omega_r} \Separator   0 \Separatorb   4 \RED{\omega_r} w^{1}_{[-3,1]} \Separator   0 \Separatorb   0 \Separatorb   12 \RED{\omega_b} \RED{\omega_r} w^{1}_{[3,-2]} \Big), \\
O_{\IND{e}}&= \tfrac{1}{8\,\RED{C}}\; F\Big( -3 w^{1}_{[3,1]} \Separator   0 \Separatorb   -6 w^{1}_{[3,-1]} w^{1}_{[3,1]} \Separator   0 \Separatorb   0 \Separatorb   36 \RED{\omega_b} \RED{\omega_r} w^{1}_{[3,1]} \Big), \\
O_{\IND{f}}&= \tfrac{1}{16\,\RED{C}}\; F\Big( w^{1}_{[9,15]} \Separator   -2 w^{2}_{[3,-9,-2]} \Separatorb   -48 w^{2}_{[1,1,-1]} \Separator   -2 \RED{\omega_b}^2 w^{1}_{[1,3]} \Separatorb   -4 \RED{\omega_b}^2 w^{1}_{[1,27]} \Separatorb   4 w^{3}_{[13,-9,-48,-4]} \Big), \\
O_{\IND{g}}&= \tfrac{1}{16\,\RED{C}}\; F\Big( -9 \RED{\omega}  \Separator   6 w^{2}_{[9,3,1]} \Separatorb   36 \RED{\omega_b}^2 \Separator   -6 \RED{\omega_b}^2 w^{1}_{[1,3]} \Separatorb   -12 w^{3}_{[4,12,6,1]} \Separatorb   12 w^{3}_{[1,3,6,2]} \Big), \\
O_{\IND{h}}&= \tfrac{1}{16\,\RED{C}}\; F\Big( 3 w^{1}_{[9,-1]} \Separator   18 \RED{\omega}  \RED{\omega'} \Separatorb   -12 w^{2}_{[6,3,7]} \Separator   0 \Separatorb   -36 \RED{\omega}  \RED{\omega'} w^{1}_{[2,1]} \Separatorb   12 w^{3}_{[-7,-3,24,10]} \Big), \\
O_{\IND{i}}&= \tfrac{1}{16\,\RED{C}}\; F\Big( -w^{1}_{[27,5]} \Separator   12 \RED{\omega_b} \RED{\omega'} \Separatorb   4 w^{2}_{[15,39,10]} \Separator   0 \Separatorb   -24 \RED{\omega_b} \RED{\omega'} w^{1}_{[1,2]} \Separatorb   4 w^{3}_{[17,-45,-48,-20]} \Big), \\
O_{\IND{j}}&= \tfrac{1}{16\,\RED{C}}\; F\Big( -\!3 w^{1}_{[1,3]} \Separator   -\tfrac{2}{\RED{\omega}} w^{3}_{[6,27,4,-5]} \Separatorb   12 w^{2}_{[5,-2,-1]} \Separator   \tfrac{6}{\RED{\omega}} \RED{\omega_r}^2 w^{2}_{[1,4,-1]} \Separatorb   4 w^{3}_{[6,27,3,2]} \Separatorb 4 w^{3}_{[-7,-3,24,10]} \Big), \\
O_{\IND{k}}&= \tfrac{1}{16\,\RED{C}}\; F\Big( w^{1}_{[21,11]} \Separator   12 \RED{\omega}  \RED{\omega'} \Separatorb   4 w^{2}_{[15,21,-4]} \Separator   0 \Separatorb   -24 \RED{\omega}  \RED{\omega_b} \RED{\omega'} \Separatorb   -4 w^{3}_{[31,33,12,20]} \Big), \\
O_{\IND{l_1}}&= \tfrac{1}{288\,\RED{C}}\; F\Big( 54 \RED{\omega} \Separator   \tfrac{3}{\RED{\omega_b}} w^{3}_{[79,66,-57,-8]} \Separatorb   -54 \RED{\omega} w^{1}_{[3,-1]} \Separator   -\tfrac{1}{\RED{\omega_b}} w^{4}_{[350,-361,114,-7,-4]} \Separatorb \nonumber\\
    &\qquad\qquad  -\tfrac{12}{\RED{\omega_b}} w^{4}_{[75,76,-33,-18,-2]} \Separatorb   -108\, \RED{\omega_r} w^{2}_{[3,4,1]} \Big), \\
O_{\IND{l_2}}&= \tfrac{1}{288\,\RED{C}}\; F\Big( -18 w^{1}_{[3,1]} \Separator   \tfrac{1}{\RED{\omega}\RED{\omega_b}} w^{4}_{[147,361,105,27,8]} \Separatorb  18 w^{1}_{[3,1]} w^{1}_{[3,5]} \Separator  -\tfrac{1}{\RED{\omega}\RED{\omega_b}} w^{2}_{[7,4,1]} w^{3}_{[14,39,-3,4]} \Separatorb \nonumber\\
    &\qquad\qquad   -\tfrac{12}{\RED{\omega_b}} w^{4}_{[21,76,21,6,2]} \Separatorb   -108\, \RED{\omega}  \RED{\omega_r} w^{1}_{[3,1]} \Big), \\
O_{\IND{m_1}}&= \tfrac{1}{32\,\RED{C}}\; F\Big( 2 w^{1}_{[9,19]} \Separator   3 \RED{\omega'} w^{1}_{[3,-11]} \Separatorb   2 w^{2}_{[-75,-150,1]} \Separator   -\RED{\omega'}^3 \Separatorb   -2 \RED{\omega'} w^{2}_{[13,-89,4]} \Separatorb   4 w^{3}_{[76,135,-24,-19]} \Big), \\
O_{\IND{m_2}}&= \tfrac{1}{32\,\RED{C}}\; F\Big( -\!6 w^{1}_{[11,3]} \Separator   w^{2}_{[-15,18,1]} \Separatorb   6 w^{2}_{[-23,38,13]} \Separator   \RED{\omega}^2 w^{1}_{[7,-5]} \Separatorb   2 \RED{\omega_b} w^{2}_{[43,54,-21]} \Separatorb   4 w^{3}_{[76,135,-24,-19]} \Big), \\
O_{\IND{n_1}}&= \tfrac{1}{12\,\RED{C}}\; F\Big( 9 w^{1}_{[3,1]} \Separator   \tfrac{\omega}{\RED{\omega_b}}  w^{2}_{[-29,14,1]} \Separatorb   -72 \RED{\omega_r}^2 \Separator   \tfrac{\RED{\omega}^2}{6 \RED{\omega_b}} w^{2}_{[56,-47,-1]} \Separatorb   -\tfrac{\RED{\omega}}{\RED{\omega_b}} w^{3}_{[-69,31,17,1]} \Separatorb   -18 w^{3}_{[9,9,-3,1]} \Big), \\
O_{\IND{n_2}}&= \tfrac{1}{12\,\RED{C}}\; F\Big( -3 w^{1}_{[3,5]} \Separator   -\tfrac{\RED{\omega'}}{3 \RED{\omega}  \RED{\omega_b}} w^{3}_{[39,31,1,1]} \Separatorb   12 w^{2}_{[9,3,4]} \Separator   -\tfrac{\RED{\omega'}^3}{6 \RED{\omega}  \RED{\omega_b}} w^{2}_{[20,23,-1]} \Separatorb \tfrac{\RED{\omega'}}{\RED{\omega_b}} w^{3}_{[39,19,13,1]} \Separatorb   -18 w^{3}_{[9,9,-3,1]} \Big), \\
O_{\IND{o_1}}&= \tfrac{1}{32\,\RED{C}}\; F\Big( 2 w^{1}_{[3,1]} \Separator   -3 R^2+\tfrac{16}{R^6} \RED{\omega}^3 \RED{\omega'}^3 \RED{\omega_r}^2-\tfrac{6}{R^2} \RED{\omega}\RED{\omega'} w^{2}_{[1,-16,1]}-3 w^{2}_{[1,0,5]} \Separatorb   -8 w^{2}_{[3,3,2]} \Separator  \nonumber\\
    &\qquad\qquad -\tfrac{12}{R^6}\, \RED{\omega}^2 \RED{\omega'}^3 \RED{\omega_r}^2 w^{2}_{[1,0,1]}+w^{3}_{[-1,30,-18,7]}-\tfrac{3}{R^2} w^{5}_{[1,-26,19,-45,18,-3]} \Separatorb \nonumber\\
    &\qquad\qquad -\tfrac{32}{R^6}\, \RED{\omega}^3 \RED{\omega'}^3 \RED{\omega_r}^2 w^{1}_{[2,1]}+6 w^{3}_{[-2,1,10,5]}-\tfrac{2}{R^2} w^{1}_{[2,-1]} w^{4}_{[5,-92,-146,-28,9]} \Separatorb   8 w^{3}_{[1,3,6,2]} \Big), \\
O_{\IND{o_2}}&= \tfrac{1}{16\,\RED{C}}\; F\Big( -3 \RED{\omega}  \Separator   \tfrac{1}{\RED{\omega}} w^{1}_{[1,-5]} w^{2}_{[3,12,5]} \Separatorb   12 \RED{\omega_b}^2 \Separator   \tfrac{1}{\RED{\omega}} w^{4}_{[1,1,12,1,13]} \Separatorb   2 w^{3}_{[2,3,54,27]} \Separatorb   4 w^{3}_{[1,3,6,2]} \Big), \\
O_{\IND{o_3}}&= \tfrac{1}{32\,\RED{C}}\; F\Big( 2 w^{1}_{[3,1]} \Separator   3 R^2-\tfrac{16}{R^6} \RED{\omega}^3 \RED{\omega'}^3 \RED{\omega_r}^2+\tfrac{6}{R^2} \RED{\omega}  \RED{\omega'} w^{2}_{[1,-16,1]}-3 w^{2}_{[1,0,5]} \Separatorb   -8 w^{2}_{[3,3,2]} \Separator \nonumber\\
    &\qquad\qquad \tfrac{12}{R^6}\, \RED{\omega}^2 \RED{\omega'}^3 \RED{\omega_r}^2 w^{2}_{[1,0,1]}+w^{3}_{[-1,30,-18,7]}+\tfrac{3}{R^2} w^{5}_{[1,-26,19,-45,18,-3]} \Separatorb  \nonumber\\
    &\qquad\qquad \tfrac{32}{R^6}\, \RED{\omega}^3 \RED{\omega'}^3 \RED{\omega_r}^2 w^{1}_{[2,1]}+6 w^{3}_{[-2,1,10,5]}+\tfrac{2}{R^2} w^{1}_{[2,-1]} w^{4}_{[5,-92,-146,-28,9]} \Separatorb   8 w^{3}_{[1,3,6,2]} \Big), \\
O_{\IND{p}}&= \tfrac{1}{8\,\RED{C}}\; F\Big( \RED{\omega_r} \Separator   -3 \RED{\omega_r}^2 \Separatorb   -2 \RED{\omega_r} w^{1}_{[6,1]} \Separator   2 \RED{\omega_r}^3 \Separatorb   2 \RED{\omega_r}^2 w^{1}_{[9,2]} \Separatorb   12 \RED{\omega_b} \RED{\omega_r} w^{1}_{[3,1]} \Big), \\
O_{\IND{q}}&= \tfrac{1}{16\,\RED{C}}\; F\Big( -w^{1}_{[3,1]} \Separator   6 \RED{\omega}  \RED{\omega_b} \Separatorb   8 w^{2}_{[3,3,1]} \Separator   -2 \RED{\omega_b}^2 w^{1}_{[1,3]} \Separatorb   -4 \RED{\omega_b} w^{2}_{[7,9,6]} \Separatorb   -4 w^{3}_{[11,21,12,4]} \Big), \\
O_{\IND{r_1}}&= \tfrac{1}{32\,\RED{C}}\; F\Big( 2 w^{1}_{[3,-1]} \Separator   -w^{2}_{[15,6,-9]} \Separatorb   -2 w^{2}_{[21,6,1]} \Separator   \RED{\omega}^2 w^{1}_{[7,-5]} \Separatorb   2 \RED{\omega} w^{2}_{[31,-1,-8]} \Separatorb   4 w^{3}_{[16,15,12,5]} \Big), \\
O_{\IND{r_2}}&= \tfrac{1}{32\,\RED{C}}\; F\Big( -2 w^{1}_{[3,5]} \Separator   -3 \RED{\omega'} w^{1}_{[13,3]} \Separatorb   2 w^{2}_{[21,-30,-23]} \Separator   -\RED{\omega'}^3 \Separatorb   -2 \RED{\omega'} w^{2}_{[1,-29,-20]} \Separatorb   4 w^{3}_{[16,15,12,5]} \Big),\\
O_{\IND{s}}&=0.
\end{align}
\noi
The $m^{2}_{x}$ are the eigenvalues of the scalar mass matrices in eqs.~\eqref{eq:mass-tree-begin}--\eqref{eq:mass-tree-end}, computed in the limit~\eqref{eq:G3211-limit-original}:\nopagebreak\vspace{-2mm}
\begin{vwcol}[widths={0.47,0.5}, sep=1cm, justify=center,rule=0pt,indent=0em]
\setcounter{equation}{53}
\begin{align} \label{eq:mass-tree-limit-begin}
m_{\IND{d}}^{2}&:= f\big(\!-\!4 \RED{\omega_r} \Separator  0 \Separatorb  24 \RED{\omega_b} \RED{\omega_r} \big), \\
m_{\IND{e}}^{2}&:= f\big(\!-\!2 w^{1}_{[3,1]} \Separator  0 \Separatorb  4 \RED{\omega_r} w^{1}_{[3,1]} \big), \\
m_{\IND{f}}^{2}&:= f\big(\!-\!2 w^{1}_{[1,2]} \Separator  2 \RED{\omega_b}^2 \Separatorb  8 \RED{\omega_b} w^{1}_{[1,2]}\big), \\
m_{\IND{g}}^{2}&:= f\big(\!-\!2 w^{1}_{[2,1]} \Separator  2 \RED{\omega_b}^2 \Separatorb  4 \RED{\omega}  w^{1}_{[2,1]}\big), \\
m_{\IND{h}}^{2}&:= f\big(\!-\!2 \RED{\omega}  \Separator  0 \Separatorb  4 \RED{\omega}  w^{1}_{[2,1]} \big), \\
m_{\IND{i}}^{2}&:= f\big(\!-\!4 \RED{\omega_b} \Separator  0 \Separatorb  8 \RED{\omega_b} w^{1}_{[1,2]} \big), \\
m_{\IND{j}}^{2}&:= f\big(\!-\!2 w^{1}_{[2,1]} \Separator  2 \RED{\omega_r}^2 \Separatorb  4 \RED{\omega}  w^{1}_{[2,1]} \big), \\
m_{\IND{k}}^{2}&:= f\big(\!-\!4 \RED{\omega}  \Separator  0 \Separatorb  8 \RED{\omega}  \RED{\omega_b} \big),\\
m_{\IND{l_1}}^{2}&:= f\big(\!-\!3 \RED{\omega}  \Separator  \tfrac{1}{2} w^{2}_{[7,-4,1]} \Separatorb  3 w^{2}_{[3,4,1]} \big), \\
m_{\IND{l_2}}^{2}&:= f\big(\!-\!w^{1}_{[3,1]} \Separator  \tfrac{1}{2} w^{2}_{[7,4,1]} \Separatorb  3 w^{2}_{[3,4,1]} \big),\\
m_{\IND{m_1}}^{2}&:= f\big(\!-\!w^{1}_{[1,3]} \Separator  \tfrac{1}{2} \RED{\omega'}^2 \Separatorb  w^{1}_{[1,3]} w^{1}_{[5,1]}\big),
\end{align}
\par\par
\begin{align}
m_{\IND{m_2}}^{2}&:= f\big(\!-\!w^{1}_{[5,1]} \Separator  \tfrac{1}{2} \RED{\omega}^2 \Separatorb  w^{1}_{[1,3]} w^{1}_{[5,1]} \big), \\
m_{\IND{n_1}}^{2}&:= f\big(\!-\!3 \RED{\omega}  \Separator  \tfrac{1}{2} \RED{\omega}^2 \Separatorb  3 \RED{\omega}  w^{1}_{[3,1]} \big), \\
m_{\IND{n_2}}^{2}&:= f\big(\!-\!w^{1}_{[3,1]} \Separator  \tfrac{1}{2} \RED{\omega'}^2 \Separatorb  3 \RED{\omega}  w^{1}_{[3,1]} \big), \\
m_{\IND{o_1}}^{2}&:= f\big(\!-\!2 \RED{\omega}  \Separator  \tfrac{3}{2} w^{2}_{[1,0,1]}\!+\!\tfrac{R^2}{2} \Separatorb  4 \RED{\omega}  w^{1}_{[2,1]} \big), \\
m_{\IND{o_2}}^{2}&:= f\big(\!-\!2 w^{1}_{[2,1]} \Separator  w^{2}_{[1,0,1]} \Separatorb  4 \RED{\omega}  w^{1}_{[2,1]} \big), \\
m_{\IND{o_3}}^{2}&:= f\big(\!-\!2 \RED{\omega}  \Separator  \tfrac{3}{2} w^{2}_{[1,0,1]}\!-\!\tfrac{R^2}{2} \Separatorb  4 \RED{\omega}  w^{1}_{[2,1]} \big), \\
m_{\IND{p}}^{2}&:= f\big(\!-\!2 \RED{\omega_r} \Separator  2 \RED{\omega_r}^2 \Separatorb  4 \RED{\omega_r} w^{1}_{[3,1]} \big), \\
m_{\IND{q}}^{2}&:= f\big(\!-\!2 \RED{\omega_b} \Separator  2 \RED{\omega_b}^2 \Separatorb  8 \RED{\omega}  \RED{\omega_b} \big), \\
m_{\IND{r_1}}^{2}&:= f\big(\!-\!\RED{\omega}  \Separator  \tfrac{1}{2} \RED{\omega}^2 \Separatorb  \RED{\omega}  w^{1}_{[5,3]} \big), \\
m_{\IND{r_2}}^{2}&:= f\big(\!-\!w^{1}_{[5,3]} \Separator  \tfrac{1}{2} \RED{\omega'}^2 \Separatorb  \RED{\omega}  w^{1}_{[5,3]} \big), \\
m_{\IND{s}}^{2}&:= 8 a_0 w^{2}_{[3,0,2]} \label{eq:mass-tree-limit-end}.
\end{align}
\end{vwcol}\vskip 2mm
\noi
The labels in the subscript of masses correspond to those defined in Table~\ref{tab:all-SM-reps} and they are further enumerated when there is more than $1$ non-vanishing eigenvalue in the relevant sector.
\vskip 2mm
\noi
The structure of the results is as anticipated:
\begin{enumerate}
\item The tree-level masses of the pseudo-Goldstone triplet and octet are proportional to $a_2$; the desired dominance of the $1$-loop contribution requires this parameter to be small.
\item
The results above (in particular, the $\Delta^{S_{INF}[\textrm{poly}]}$ and $\Delta^{S[\textrm{log}]}$ terms) are written in a simplified form corresponding to the limit~\eqref{eq:G3211-limit-original}\footnote{Since the largest mass matrix one has to deal with here is `only' $4\times 4$ it should nevertheless be possible to write down the results also in the situation when the limit~\eqref{eq:G3211-limit-original} is not imposed. However, as we plan to use these mainly for the sake of a future numerical study where the diagonalization of such structures is trivial, there is no real reason to do that.}. Note, however, that in full generality $a_2$ is strongly correlated to $|\sigma|^2$ and $\tau$ (which has to be well below $M_{\rm Pl}$ in order to retain the perturbative regime). At one loop, it is sufficient to insert the tree-level expression~\eqref{eq:tau-tree-level} for $\tau$ into the $1$-loop $\Delta$-terms.
\item In the limit~\eqref{eq:G3211-limit-original}, the largest blocks of the scalar mass matrix evaluated in the vacuum are $2\times 2$ and the only one which is not diagonalized trivially is $M^{2}_{\IND{o}}$, see~\eqref{eq:Mo}.  This is where the square-root \eqref{eq:Rsquared} may arise and it does so in $\Delta_b^{S_{INF}[\textrm{poly}]}$  due to the non-commutativity of the relevant $M^{2}_{\IND{o}}$ block with the corresponding part of the derivatives of the field dependent mass matrix (requiring diagonalization
    and normalization of the eigenmodes as suggested by eqs.~\eqref{eq:general-nested-commutator}--\eqref{eq:definition-Mb}).
    The purely polynomial terms emerge solely from the degeneracy of the scalar spectrum in the SM symmetry limit along the lines described in Sect.~\ref{sect:methods}.
    Furthermore, rational functions including the $R$-factor also appear in the coefficients $T_{x}$ and $O_{x}$ of logarithmic contributions $\Delta_a^{S[\textrm{log}]}$ and $\Delta_b^{S[\textrm{log}]}$ (in particular, in terms with $x\in\{o_1,o_3\}$).

\item The logarithmic terms in $\Delta^{G[\textrm{log}]}$ and $\Delta^{S[\textrm{log}]}$ have arguments of the form $m^2/\mu_r^2$, where $m$ is a mass of a physical particle (gauge boson or scalar) and $\mu_r$ is the renormalization scale. Only terms corresponding to particles with non-vanishing mass turn out to be present, i.e.,~there are no IR divergences in the results (for $\Delta^{S[\textrm{log}]}$, the limit~\eqref{eq:G3211-limit-original} was used; in further limits of the regular- and flipped-$SU(5)$ case IR divergences do appear, see comments in Sections~\ref{sect:mass-shell} and~\ref{sec:checks-limits}). Out of the list of massive particles, all have terms present except for two exceptions: the gauge singlet contribution in $\Delta^{G[\textrm{log}]}$ and the massive singlet present in $M^2_{\IND{s}}$ is missing from $\Delta^{S[\textrm{log}]}$, i.e.~$T_{\IND{s}}=O_{\IND{s}}=0$. Note that the use of limit~\eqref{eq:G3211-limit-original} gives an extra would-be Goldstone singlet in $M^2_{\IND{s}}$ (because the symmetry is increased), while states $\IND{a}$, $\IND{b}$ and another singlet in $M_{\IND{s}}^2$ become massless due to their pseudo-Goldstone nature; the limit gives $22$ massive fields instead of $26$ which, given the absence of $m_{\IND{s}}^2$, yields $21$ contributions to $\Delta^{S[\textrm{log}]}$, see eq.~(\ref{eq:logs}).
\end{enumerate}

\section{Evaluating the nested commutators -- further details\label{app:nested-commutator-evaluation}}
In this Appendix we intend to clarify several technical points related to the evaluation of the (traces of) infinite series of nested commutators~\eqref{eq:nested-commutator-series}. Its direct term-by term evaluation is in general rather complicated, mainly due to the fact that there is no general connection among $\mathbf{A}$, $\mathbf{A}_a$ and $\mathbf{A}_b$ (which correspond to different orders in the Taylor expansion of the field-dependent mass matrices $\mathbf{M}^{2}_{\mathbf{S,G}}(\Phi)$ around the tree-level vacuum) and, on top of that, the expected $a\leftrightarrow b$ symmetry is not apparent on the RHS of formula~\eqref{eq:nested-commutator-series}.
The method described below and used to deal with this issue yields not only an interesting analytic insight but it also provides a relatively simple procedure for the analytic evaluation that becomes almost trivial in case of a numeric treatment.

The argument goes as follows:
any of the $n\times n$ matrices $\mathbf{A}$, $\mathbf{A}_a$ and $\mathbf{A}_b$ can be viewed as an element of the space of endomorphisms $\mathrm{End}(\mathbb{R}^n)$. Since $\mathbf{A}$ is proportional to the mass matrix, it is Hermitian and hence diagonalizable; its spectral decomposition looks like
$
\mathbf{A}=\sum_{i=1}^{n} \lambda_i\,\mathbf{v}_i \mathbf{v}_i^\dagger
$
with dimensionless eigenvalues $\lambda_i=m^2_i/\mu_r^2$ corresponding to the normalized eigenvectors $\mathbf{v}_i$ forming an orthonormal basis in $\mathbb{R}^{n}$, $\mathbf{v}_i^\dagger \mathbf{v}_j=\delta_{ij}$.
Consequently, the matrices $\mathbf{v}_i \mathbf{v}_j^\dagger$ form an orthonormal basis of the $n^2$-dimensional space of $\mathrm{End}(\mathbb{R}^n)$, $\Tr\left[\mathbf{v}_i \mathbf{v}_j^\dagger\;(\mathbf{v}_k \mathbf{v}_l^\dagger)^\dagger\right]=\delta_{ik}\delta_{jl}$.
Notice that the commutator with $\mathbf{A}$, defined as $\mathrm{Ad}_\mathbf{A} (\mathbf{B})\coloneqq[\mathbf{A},\mathbf{B}]$,
is a linear operator on the space of matrices $\mathrm{End}(\mathbb{R}^n)$ and, hence $\mathrm{Ad}_{\mathbf{A}}\in\mathrm{End}^{2}(\mathbb{R}^n)$. Crucially, the set of matrices $\{\mathbf{v}_i \mathbf{v}_j^\dagger\}$ forms a complete eigenbasis for the $\mathrm{Ad}_{\mathbf{A}}$ operator, with respective real eigenvalues $\lambda_i-\lambda_j$. The nested commutators with $\mathbf{A}$ are then easily evaluated on these basis elements:
\begin{align}
\underbrace{[\mathbf{A},[\mathbf{A},\ldots[\mathbf{A},\mathbf{v}_i\mathbf{v}_j^\dagger]]]}_{k-1\;\;\textrm{times}}=\mathrm{Ad}_{\mathbf{A}}^{k-1}\, (\mathbf{v}_{i}\mathbf{v}_j^\dagger )=(\lambda_i-\lambda_j)^{k-1} \, \mathbf{v}_i\mathbf{v}_j^\dagger.
\end{align}
Expanding now $\mathbf{A}_a$ and $\mathbf{A}_b$ in the $\{\mathbf{v}_i\mathbf{v}_j^\dagger\}$ eigenbasis as
$
\mathbf{A}_a=\sum_{i,j=1}^{n}M^a_{ij}\, \mathbf{v}_i\mathbf{v}_j^\dagger$, $
\mathbf{A}_b=\sum_{i,j=1}^{n}M^b_{ij}\, \mathbf{v}_i\mathbf{v}_j^\dagger,
$
with coefficients $M^a_{ij}$ and $M^b_{ij}$ (which can be alternatively viewed as the matrix elements of $\mathbf{A}_a$ and $\mathbf{A}_b$, cf. eqs.~\eqref{eq:definition-Ma}--\eqref{eq:definition-Mb}, rewritten in the eigenbasis of $\mathbf{A}$), one arrives at
\begin{align}
\mathbf{\Upsilon}(\mathbf{A},\mathbf{A}_a,\mathbf{A}_b)&=
\sum_{p,q=1}^{n}\sum_{i,j=1}^{n} M^a_{pq} M^{b}_{ij}\;
\sum_{m=1}^{\infty}(-1)^{m+1}\,\frac{1}{m}\sum_{k=1}^{m}\binom{m}{k}\left\{\mathbf{A},\mathbf{v}_p\mathbf{v}_q^\dagger\right\}\mathrm{Ad}_{\mathbf{A}}^{k-1}(\mathbf{v}_i\mathbf{v}_j^\dagger)\;(\mathbf{A}-\mathbb{1})^{m-k}
\nn\\
&=\sum_{p,q=1}^{n}\sum_{i,j=1}^{n} M^a_{pq} M^{b}_{ij}\;
\mathbf{v}_p\mathbf{v}_q^\dagger\mathbf{v}_i\mathbf{v}_j^\dagger\;(\lambda_p+\lambda_q)
\times \left\{
\begin{array}{cl}
\frac{\log\lambda_i-\log\lambda_j}{\lambda_i-\lambda_j}
& \textrm{ for } \lambda_i \neq \lambda_j \\
\frac{1}{\lambda_j} & \textrm{ for } \lambda_i = \lambda_j
\end{array}
\right.
.
\end{align}
In case of degenerate eigenvalues, i.e.~$\lambda_i=\lambda_j$, only the $k=1$ term gives a non-zero contribution.
With this at hand the trace of $\mathbf{\Upsilon}$ is obtained readily:
\begin{align}
\Tr\mathbf{\Upsilon}(\mathbf{A},\mathbf{A}_a,\mathbf{A}_b)&=\sum_{i,j;\;\lambda_i\neq \lambda_j}\!\!M^{a}_{ji}M^{b}_{ij}\,\frac{\lambda_i+\lambda_j}{\lambda_i-\lambda_j}\,\log\frac{\lambda_i}{\lambda_j} + \sum_{i,j;\;\lambda_i=\lambda_j}\!\!2M^{a}_{ji}M^{b}_{ij},\label{eq:general-nested-commutator-final}
\end{align}
and one arrives at the formula~\eqref{eq:general-nested-commutator}. Note also that the $a\leftrightarrow b$ symmetry becomes manifestly apparent in this form.
\vskip 2mm
\noindent
Eq.~\eqref{eq:general-nested-commutator-final} gives some interesting insights into the contribution of the nested commutator series to the masses:
\begin{enumerate}
\item Notice that there are two types of contributions in equation~\eqref{eq:general-nested-commutator-final}, polynomial and logarithmic. The arguments of the logs are the eigenvalues of $\mathbf{A}=\mathbf{M}_{\mathbf{S,G}}^2/\mu_r^2$, so they give rise to exactly the same types of logs as the usual matrix logarithms in the rest of eq.~\eqref{eq:1loop-masses} and thus analytic results are again subject to the same limitation that is due to the diagonalization procedure, which cannot be worked out analytically for sufficiently complicated matter content. That ceases to pose a problem in the numerical treatment. From this perspective, the result for $\mathbf{\Upsilon}$ is as good as could be hoped for.
\item The polynomial contribution is due to the degeneracy of the eigenvalues of $\mathbf{A}$, which is generally the case of any non-Abelian gauge theory with non-trivial matter content. More symmetry implies more degeneracy, and thus more terms of the polynomial type. If one turns off certain VEVs in a spontaneously broken scenario, the degeneracy of the masses increases and further logarithmic contributions transform into polynomials in that limit. Some explicit examples of such behaviour can be found in Section~\ref{sec:checks-limits}.
\item Notice that the result is invariant under the rescaling $\lambda_k\to \kappa\lambda_k$ with a common factor $\kappa$ for all $k$, and thus does not explicitly depend on the renormalization scale $\mu_r$ (unlike other terms in the 1 loop correction to the mass).
\item The form of the result~\eqref{eq:general-nested-commutator-final} admits for a simple isolation of the spurious IR divergences due to the vanishing log arguments corresponding to Goldstone bosons, cf. Sect.~\ref{sect:mass-shell}.
\end{enumerate}


\end{document}